  \documentclass[%
    reprint,
    superscriptaddress,
    preprintnumbers,
    amsmath,amssymb,
    aps,
    prb,
    floatfix,
    ]{revtex4-2}
    \usepackage[shortlabels]{enumitem}% Include figure files
    \usepackage{graphicx}% Include figure files
    \usepackage{dcolumn}% Align table columns on decimal point
    \usepackage{tabularx}
    \usepackage{booktabs}

    \usepackage{bm}
    \usepackage{xstring}
    \usepackage{physics}
    \usepackage[usenames,dvipsnames]{xcolor}
    \usepackage{hyperref}
    \usepackage{multirow}
    \usepackage{array}

     \usepackage{soul}

    \usepackage{mathtools}

    \usepackage{soul}

  \usepackage{colortbl}
  \definecolor{Gray}{gray}{0.85}
  \newcolumntype{H}{>{\columncolor{Gray}}c}
  \newcolumntype{H}{>{\columncolor{OliveGreen!15!white}}c}

    \newcommand{\beqref}[1]{Eq.~\eqref{#1}}

  \usepackage[defaultcolor=OliveGreen]{changes}

  \makeatletter
  \AddToHook{cmd/added/before}{\def\Changes@AuthorColor{OliveGreen}}
  \AddToHook{cmd/deleted/before}{\def\Changes@AuthorColor{lightgray}}
  \AddToHook{cmd/replaced/before}{\def\Changes@AuthorColor{OliveGreen}}
  \makeatother

  \let\oldcitep\citep
  \renewcommand{\citep}[1]{\mbox{\oldcitep{#1}}}
  \let\oldcitet\citet
  \renewcommand{\citet}[1]{\mbox{\oldcitet{#1}}}
  \let\oldcite\cite
  \renewcommand{\cite}[1]{\mbox{\oldcite{#1}}}

  \newcommand{\w}{\omega}
  
  \newcommand{\e}{\epsilon}

  \def\E{\mathcal{E}}
  
  \def\Ea{{\mathcal{E}_{\alpha}}}

    \usepackage{sansmath}

\AtBeginDocument{\let\origref\ref\renewcommand{\ref}[1]{\mbox{\origref{#1}}}}
\begin{document}

  \title{Plasmon-phonon coupling from frequency-dependent density-functional perturbation theory}

  \author{Guglielmo Marchese}
  \email{gmarchese@icmab.es}
  \affiliation{Institut de Ciència de Materials de Barcelona (ICMAB-CSIC), Campus UAB, 08193 Bellaterra, Spain}

  \author{Cyrus E. Dreyer}
  \affiliation{Center for Computational Quantum Physics, Flatiron Institute,162 5th Avenue, New York, New York 10010, USA}
  \affiliation{Department of Physics and Astronomy, Stony Brook University, Stony Brook, New York 11794-3800, USA}

  \author{Massimiliano Stengel}
  % \email{mstengel@icmab.es}
  \affiliation{Institut de Ciència de Materials de Barcelona (ICMAB-CSIC), Campus UAB, 08193 Bellaterra, Spain}
  \affiliation{ICREA - Institució Catalana de Recerca i Estudis Avançats, 08010 Barcelona, Spain}

  \date{\today}% It is always \today, today,

  \begin{abstract}
  While doped semiconductors are a pillar of modern technology, their 
  theoretical characterization remains challenging. Standing between the 
  insulating and metallic asymptotic regimes, the intermediate doping regime--whose manipulation underlies applications ranging from transistors and solar 
  cells to light-emitting diodes--lies beyond the reach 
  of standard approximations. In particular, the electronic density and the 
  lattice cannot be regarded as uncoupled systems, and hybrid excitations 
  characterize the low-doping spectrum where the plasma frequency is on the order of the lattice vibration frequencies. In this scenario, the response functions of the doped crystal and their frequency dependence must in general be retained, or their neglect justified, to capture the interplay of coupled excitations.
  We present an efficient first-principles procedure to describe plasmon-phonon 
  coupling at long wavelength ($\mathbf q\to0$) based on Legendre transformations, which recasts the interaction as a 
  frequency-dependent eigenvalue problem. Exploiting recent advances in 
  Time-Dependent Density Functional Perturbation Theory (TD-DFPT) methods,
  we verify the generalized acoustic sum rule, which expresses momentum conservation, achieving its verification in metals at 
  finite complex frequency with the recently developed traveling 
  pseudopotential correction. We further present a fast interpolation procedure 
  to calculate entire spectra at arbitrary doping with benchmarked accuracy.
  We validate our approach against experimental data for BaSnO$_3$, 
  a perovskite transparent conductor, finding good agreement with the infrared dielectric function and with the plasma frequency measured at known carrier density.
  \end{abstract}

  \maketitle

  \renewcommand{\figurename}{}
  \renewcommand{\thefigure}{\textbf{Fig.~\arabic{figure}}}

  % \onecolumngrid
  % \tableofcontents
  % \clearpage
  \twocolumngrid

  % \section*{Ask Max}
  % \begin{itemize}
    
  % \end{itemize}

  \section{Introduction}
  Metallic crystals can support collective charge oscillations of the free
  carriers that are commonly known as plasmons~\cite{pines_collective_1952}.  In doped semiconductors, their
  characteristic frequency ($\omega_p$) often overlaps with optical phonon
  frequencies ($\omega_{\rm ph}$), leading to hybrid plasmon-phonon modes~\cite{varga_coupling_1965,singwi_interaction_1966}.  
  %
 % \ms{This sentence feels truncated (aren't there other reasons why we are interested in these modes, apart from mobility?), missing ref}
  Such coupled modes, first observed in the Raman spectra of $n$-type
  GaAs~\cite{mooradian_observation_1966,mooradian_polarization_1967},
  reshape the low-energy excitation spectrum and the screening of the
  long-range Coulomb interaction: they limit the carrier
  mobility~\cite{kim_electron_1978,rajapitamahuni_plasmonphonon_2022},
  renormalize the quasiparticles and their polaronic
  satellites~\cite{verdi_origin_2017,caruso_electron-plasmon_2018,lihm_plasmon-phonon_2024},
  and dominate the infrared response of doped polar
  crystals~\cite{streyer_engineering_2016}.
  As a consequence, a predictive understanding of plasmon-phonon
  resonances is critical to several frontier technologies, from the design
  of transparent conductors~\cite{ginley_transparent_2000} to infrared
  plasmonics and polaritonics~\cite{law_towards_2013,caldwell_low-loss_2014},
  where hybrid modes extend the Reststrahlen
  band~\cite{streyer_engineering_2016} and enable ultrafast optical
  switching~\cite{dunkelberger_active_2018}.
  The fundamental and practical interest in such fields is rapidly growing,
  with the advent of hyperbolic materials, in which hybrid plasmon-phonon
  polaritons can be tuned by gating~\cite{dai_graphene_2015}, enabling
  previously inaccessible regimes of light-matter
  interaction~\cite{basov_polaritonic_2025,hu_transport_2026,venturi_visible-frequency_2024}.

  Depending on the ratio $\omega_p/\omega_{\rm ph}$, the low-energy spectrum
  falls into three regimes~\cite{verdi_origin_2017,lihm_plasmon-phonon_2024}, in
  which different levels of approximation hold and the theoretical challenges
  change accordingly.
  In the limit of small carrier density ($\omega_p \ll \omega_{\rm ph}$),
the plasmon lies well below the optical phonons, while the interband
electronic excitations lie far above both, typically at the scale of the
band gap, well outside the low-energy spectrum.
  In this anti-adiabatic regime the plasma
  oscillation is frozen on the time scale of the lattice vibrations, which means that the dynamics of free carriers has negligible impact on phonon frequencies.
  Thus, from the lattice-dynamical perspective, one can safely assume an insulating crystal, whose main physical signature is the frequency splitting between longitudinal (LO) and transverse (TO) optical modes~\cite{born_dynamical_1954,cochran_dielectric_1962}.
%  The adiabatic approximation then holds for the bound electrons, and the 
 % Lattice dynamics---including the long-range dipolar interactions responsible for the
  %---is
 % fully determined by static response functions: the interatomic force constants,
 % the Born effective charges, and the high-frequency dielectric  tensor
  The microscopic formulation of this 
  picture~\cite{sham_electronic_1969,pick_microscopic_1970,vogl_microscopic_1976}
  and its implementation within density-functional perturbation theory
  (DFPT)~\cite{gonze_dynamical_1997,baroni_phonons_2001} are now the gold standard
  for polar insulators.
  In the opposite, metallic limit ($\omega_p \gg \omega_{\rm ph}$), the Drude
  oscillation of the carriers in partially occupied
  bands is much faster than the lattice and
  fully screens the macroscopic field associated with polar phonons: the electrostatic
  contribution to the LO phonon energy is suppressed, and the LO-TO splitting closes. 
  This regime is also well captured by the static force constants, whose generalization to metals is now part of the standard DFPT toolkit~\cite{de_gironcoli_lattice_1995,baroni_phonons_2001}.
  % \g{quest'accenno mi sembra sbagliato cosi' come e': Dynamical effects such as Kohn-anomalies or Fermi surface contributions, are due to short-range electron-phonon coupling \cite{calandra_adiabatic_2010} and are not mediated to the long-range Coulomb.}

  At intermediate carrier densities ($\omega_p \sim \omega_{\rm ph}$), neither picture 
 % \ms{simplifying assumption} 
  applies. The carriers are too slow to follow the lattice adiabatically,
  yet too fast to be neglected, and the partially screened long-range Coulomb
  interaction mixes the plasmon with the polar phonons.  
  Both excitations, and their coupling, then require a dynamical, nonadiabatic
treatment. This is a demanding task for theoretical modeling; for decades, the
theory of coupled plasmon-phonon modes has relied on phenomenological approaches,
often based on adjustable parameters or on simplified descriptions of the electronic
and phonon dispersions~\cite{yokota_coupling_1961,varga_coupling_1965,
singwi_interaction_1966,falter_effect_1993,bauer_impact_2009,
hwang_plasmon-phonon_2010,krsnik_importance_2022}.
  
  On the first-principles front, 
 the study of plasmons has made remarkable progress since the early studies of the uniform electron gas ~\cite{bohm_collective_1953,pines_collective_1952,lindhard_properties_1954}. 
 Calculations on real metals 
  are now routinely performed 
  either within the
  random-phase approximation (RPA)~\cite{maddocks_ab_1994} or at the
  time-dependent density-functional theory (TDDFT)~\cite{quong_first-principles_1993} level. Efficient implementations of the latter
  in the frequency domain~\cite{timrov_electron_2013,botti_time-dependent_2007,varas_quantum_2016}
  or in real time~\cite{bertsch_real-space_2000} are now available.
  Meanwhile, first-principles approaches to  nonadiabatic phonon frequencies have seen remarkable progress over the past decades~\cite{lazzeri_nonadiabatic_2006,calandra_adiabatic_2010}.
  Most of these studies, however, relied on the variational approximation of Ref.~\cite{calandra_adiabatic_2010}, where  
  electronic screening is treated statically.
  While appropriate in several contexts, such an approximation is unfit for our scope, as it does not capture the dynamical charge fluctuations that underpin plasmon resonances.

  Serious attempts at treating the coupled problem from first principles appeared
  only recently. In Ref.~\cite{lihm_plasmon-phonon_2024}, plasmons are represented as quasiparticles whose frequency and lifetime are fitted from the calculated RPA spectra.
  They are subsequently coupled to phonons via an extended dynamical matrix approach, in close analogy with the (quasi-)adiabatic treatment of magnon-phonon hybrids~\cite{ren_adiabatic_2024}.
 Recent work on layered
  systems~\cite{macheda_ab_2024} follows a similar strategy for the interlayer plasmon-phonon interaction, 
  with the plasmon-phonon coupling treated via an electrostatic
  model~\cite{dolgov_admissible_1981}.
 % (STILL NEED TO TIGHTEN THE FOLLOWING PART...)
  %
  While they constitute a very significant step forward, however, 
  both schemes rely on simplifying assumptions where 
  plasmons and phonons are treated at different levels of approximation.
  As such, their performance with regard to exact sum rules and 
conservation laws~\cite{dreyer_nonadiabatic_2022,stengel_rototranslational_2026} relating 
the nuclear and electronic subsystems is unclear.
  Most importantly, a general prescription for treating systems where such 
  simplifying assumptions are not applicable is currently missing.

  Here we fill this gap by presenting a fully \textit{ab initio}
  approach to coupled plasmon-phonon resonances.
  Starting from an extended Ehrenfest action where the macroscopic vector
  potential and the ionic positions both appear as dynamical variables, 
  we derive the equations of motion for their coupled evolution in frequency space.
  We work in the long-wavelength limit $\mathbf q\to0$, where the macroscopic field is uniform; the plasmon dispersion and Landau damping at finite $\mathbf q$ are left to future work.
The resonances are then determined by setting to zero the determinant of the frequency-dependent Hessian matrix in the extended space, whose subblocks correspond to the dielectric function, Born effective charges and interatomic force constants.
 In this way, the free carriers and lattice degrees of freedom are treated 
 on the same footing, yielding the plasmon and phonon interplay at the time-dependent DFPT (TD-DFPT) level of theory
once the appropriate electrical (open-circuit) and mechanical (relaxed-ion) boundary conditions  
are imposed~\cite{resta_polarization_2006,stengel_electric_2009}.

 In practice, the relevant material coefficients are computed via a dynamical generalization of DFPT for metals that 
 builds on some recent
developments in this area~\cite{dreyer_nonadiabatic_2022,binci_first-principles_2021,
  marchese_born_2024,paper2}. 
Crucially, our calculations are always
performed at vanishing macroscopic electric field and clamped ions, while the boundary conditions are dealt with at the post-processing stage.
%
%To that end, we build on 
Thanks to the Legendre-transform framework of Ref.~\cite{royo_dynamical_2026},
%to switch from transverse/longitudinal and clamped-ions/relaxed-ions conditions for an arbitrary propagation direction 
the complete low-energy spectrum of the system (including transverse and longitudinal modes) can be obtained via simple linear-algebra operations on the extended Hessian.
%
%  Therefore, unlike in Ref.~\cite{stengel_electric_2009}, the
%  boundary conditions need not be imposed within the
%  self-consistent TD-DFPT cycle.
 % and can be changed without repeating the
 % calculation or losing accuracy.
  %
  The frequency-independent matrix of Lihm and
Park~\cite{lihm_plasmon-phonon_2024} is recovered, at $\mathbf q\to0$, as a well-defined adiabatic approximation of our method. 
The latter consists of neglecting the doping and frequency dependence of the bound-charge susceptibility, force constants, and Born effective charges, which is very accurate in situations where the interband gap is much larger than the phonon energies.
%  $\mathbb{S}(\w)$: approximating the carrier susceptibility by a
%  plasmon pole allows its frequency dependence to be absorbed into an
%  auxiliary plasmon coordinate, and taking the remaining ingredients from
%  the undoped crystal in the static limit yields their model. Our
%  formulation thus lifts these approximations within a single
%  self-consistent framework.
  
  Our unified approach to plasmons and phonons provides an ideal platform for testing exact sum rules and illustrating their physical significance. 
  Of particular note, we provide   a direct numerical verification 
  of the generalized acoustic sum rule~\cite{dreyer_nonadiabatic_2022} at finite frequency in
  conducting systems, demonstrating the necessity of the traveling-pseudopotential
  correction of Ref.~\cite{stengel_rototranslational_2026}. 
  Crucially, we show that this sum rule enables a coupling between acoustic phonons 
  and plasmons that does not vanish in the long-wavelength limit.
  Although numerically small, such a coupling is required for the internal consistency of the theory: 
     it recovers the textbook formula for the ionic correction to the plasma frequency, 
where the bare electron mass is replaced by the reduced mass of the coupled electron-ion system.
  %
  % \st{Finally, the fully self-consistent method allows us to assess
  % controlled approximations.
  % We devise an interpolation strategy %(Sec.~\ref{sec:interp}) 
  % that
  % covers the whole frequency-doping parameter space at the cost of few
  % explicit finite-frequency calculations. Its accuracy is monitored by the
  % generalized acoustic sum rule, whose violation provides a strict consistency test
  % indicating where additional TD-DFPT calculations are required.} \ms{(Is there any technical innovation in your interpolation method? If not, I would not mention it here.)}
  %\note{application}
%
  We apply this framework to BaSnO$_3$ (BSO), a cubic perovskite oxide that,
  upon La doping, combines high electron mobility with optical transparency,
  making it a promising platform for transparent conductors, plasmonic
  devices, and oxide
  electronics~\cite{luo_high_2012,kim_high_2012,kim_physical_2012,
  ismail-beigi_alkaline_2015,monserrat_phonon-assisted_2018}. 
  %As the carrier
  %density is varied across the insulator-to-metal transition, 
  As a function of the carrier density, the
  low-energy spectrum of BSO spans all three regimes of the plasmon-phonon
  interplay---unscreened, strongly coupled, and fully screened---making it an
  ideal testbed. 
  Our results agree well with the measured far-infrared
  dielectric response~\cite{stanislavchuk_electronic_2012} and with the
  plasma frequency measured at known carrier density~\cite{james_allen_conduction_2016}.

  %\note{paper structure}
  This paper is organized as follows. Section~\ref{sec:theory} introduces the
  dielectric response functions and the electrical boundary conditions
  (Sec.~\ref{sec:diel}), derives the TD-DFPT action of the coupled system
  (Sec.~\ref{sec:action}), and obtains from it the plasmon-phonon equations
  of motion (Sec.~\ref{sec:extD}). Sections~\ref{sec:sumrule}--\ref{sec:pla_renorm}
  apply this framework to the generalized acoustic sum rule, the pseudo-static approximation, a minimal model, and the
  renormalization of the plasmon by the lattice. Section~\ref{sec:method}
  discusses the implementation and the approximation strategies, and
  Sec.~\ref{sec:appl_res} presents the low-energy spectrum of BSO and its
  comparison with experiment. Section~\ref{sec:concl} summarizes our
  conclusions.

  \section{Theory}\label{sec:theory}

  \subsection{%\st{Electric response}
  Longitudinal dielectric function}\label{sec:diel}

  The properties of plasmons, like those of any electromagnetic excitation of a
material, are encoded in the momentum- and frequency-dependent dielectric tensor~\cite{born_dynamical_1954,pines_collective_1952,melrose_electromagnetic_1991}.
Because plasmons are longitudinal oscillations of the electronic charge density,
they are naturally associated with
%\ms{propagation is governed by} 
%long-wavelength behavior is conveniently accessed at the Brillouin-zone center
%by imposing 
open-circuit electrical boundary conditions (EBC)~\cite{resta_polarization_2006}, i.e., a vanishing
electric displacement field along the propagation direction~\cite{stengel_electric_2009}.
Under such conditions, the internal electric field acts as a depolarizing response to
the local accumulation or depletion of free carriers.
The plasmon dynamics is then governed by the inverse dielectric function, whose
macroscopic, long-wavelength limit is~\cite{nozieres_electron_1959}
  \begin{align}
      \epsilon^{-1}(\w) \equiv \frac{1}{\epsilon_{\rm L}(\w)} 
      = \lim_{\mathbf{q}\to\mathbf{0}}
      \frac{1}{\hat{\mathbf{q}}\cdot
      {\bm\epsilon}(\mathbf{q},\w)\cdot\hat{\mathbf{q}}}.
      \label{eq:inv_diel}
  \end{align}
  The poles of $\epsilon^{-1}(\w)$, including plasmons and polar longitudinal optical (LO) phonons, are determined by
%
%  This yields 
  the following resonance condition,
  %on the dielectric function:
  \begin{align}
     {\rm Re} \,\,  \epsilon_{\rm L}(\w) = 0.
      \label{eq:diel=0}
  \end{align}
Thus, all the relevant information is contained in the zone-center dielectric tensor $\epsilon_{\alpha\beta}(\w)$ \cite{melrose_electromagnetic_1991,keldysh_dielectric_1989}.
 
 %\ms{(YOU HAVEN'T DEFINED THE ZONE CENTER DIELECTRIC TENSOR YET)}
%\g{intendi di aggiungere altro, o hai gia' risolto tu cosi'?}

  The dielectric tensor receives two distinct
  contributions: an electronic part $\chi^{\rm el}_{\alpha\beta}(\w)$,
  calculated with the ions clamped at their equilibrium positions, and a
  vibrational part $\chi^{\rm vib}_{\alpha\beta}(\w)$ arising from the
  ionic motion~\cite{born_dynamical_1954},
  \begin{align}
    \epsilon_{\alpha\beta}(\w) = \delta_{\alpha\beta} + 4\pi\chi^{\rm el}_{\alpha\beta}(\w)
    + 4\pi \chi^{\rm vib}_{\alpha\beta}(\w).
    \label{eq:diel_tot}
  \end{align}
  % Equation~\eqref{eq:diel_tot} is routinely evaluated in the static limit 
  % for insulating systems. However, retaining its full frequency dependence and 
  % treating $\chi^{\rm el}_{\alpha\beta}(\w)$ properly extends its 
  % validity to metals and doped semiconductors.
  The electronic contribution, in non-magnetic systems, consists of an interband ``bound-charge'' part
  $\chi^{\rm el,bound}_{\alpha\beta}(\w)$, arising from transitions between
  occupied and empty bands, which is also present in insulators, and
  a Drude contribution, present only in metals and proportional to the Drude weight $D_{\alpha\beta}$,
  \begin{align}
    { \chi}^{\rm el}_{\alpha\beta}(\w) = - \frac{{D}_{\alpha\beta}}{\pi \w^2}
    + { \chi}_{\alpha\beta}^{\rm el,bound}(\w).
    \label{eq:chi_el}
  \end{align}
 % In the metallic regime, the plasmon is well described by the 
 % electric contribution alone and can be obtained by the
%  longitudinal clamped-ion dielectric function 
%  \begin{align}
 %     \e^{\rm el}(\omega) = 1 + 4\pi\lim_{\mathbf{q}\to\mathbf{0}}
  %    \hat{\mathbf{q}}\cdot{\bm\chi}^{\rm el}(\mathbf{q},\omega)
   %   \cdot\hat{\mathbf{q}}.
 % \end{align}
 %
  The Drude term diverges at low frequencies
  and is
  responsible for the qualitatively different 
   behavior of conducting systems, since the other contributions, ${ \chi}^{\rm el,bound}_{\alpha\beta}(\w)$ and $\chi^{\rm vib}_{\alpha\beta}(\w)$, are regular in the static limit. Beyond its role in the dielectric response, the
  Drude weight is regarded as a defining geometric property of the metallic state,  characterizing the degree of
  electronic delocalization in the system~\cite{resta_drude_2018}. %quantifying the electronic density available for DC conductivity. 
  In a Kohn--Sham or independent-particle framework 
  it can be related to the Fermi-sea integral of the inverse effective mass tensor via~\cite{kohn_theory_1964}
  \begin{align}
    D_{\alpha\beta} = \frac{\pi e^2 s_d }{\Omega\hbar^2}
    \int_{\rm BZ}[d{\bf k}]  \sum_l f_{\mathbf{k}l}
    \frac{\partial^2\e_{\mathbf{k}l}}
    {\partial k_\alpha\,\partial k_\beta},
    % \nonumber \\
    % {\color{red}
    % -\frac{\pi e^2 s_d }{\Omega\hbar^2}
    % \int_{\rm BZ}[d{\bf k}]  \sum_l f'_{\mathbf{k}l}
    % \frac{\partial\e_{\mathbf{k}l}}
    % {\partial k_\alpha}\frac{\partial\e_{\mathbf{k}l}}
    % {\partial k_\beta},
    % }
    \label{eq:drude}
  \end{align}
  where $s_d$ is the spin degeneracy, $\Omega$ is the unit-cell volume, $f_{\mathbf{k}l}$ is the Fermi occupation evaluated at the band energy $\e_{\mathbf{k}l}$, and $[d{\bf k}]=\Omega/(2\pi)^3 d^3k$ is the reciprocal-space differential normalized by the Brillouin-zone ($\rm BZ$) volume. In the free-electron model 
  the inverse effective mass reduces to a Kronecker delta, constant over the parabolic band;
%  where all electrons participate in conduction, 
  the Drude weight is then proportional to the number of electrons  $N^{\rm el}$~\cite{ashcroft_solid_1976}
  \begin{align}
    D_{\alpha\beta} =  \frac{\pi e^2}{ \Omega} \frac{N^{\rm el} }{ m_e} \delta_{\alpha\beta}.
    \label{eq:drude_jellium}
  \end{align}
  In the metallic regime, the plasmon is captured by the electronic
contribution alone, i.e., by the longitudinal clamped-ion dielectric function
      $\e^{\rm el}(\omega) = 1 + 4\pi  
      \hat{\mathbf{q}}\cdot{\bm\chi}^{\rm el}
       (\omega)
      \cdot\hat{\mathbf{q}}$.
  
 The vibrational contribution to the dielectric response is~\cite{maradudin_theory_1971,gonze_dynamical_1997}
  \begin{align}
    \chi^{\rm vib}_{\alpha\beta}(\w) = \frac{e^2}{\Omega}
    Z_{s\alpha\alpha'}(-\w)
    \bigl[\mathbf{C}^0(\w) - \mathbf{M}\w^2 \bigr]^{-1}_{ss'\alpha'\beta'}
    Z_{s'\beta\beta'}(\w),
    \label{eq:chi_vib}
  \end{align}
  where $s,s'$ run over the unit-cell ion sites, $M_{ss'}= m_s \delta_{ss'}$ is the ionic mass
  matrix, $Z_{s\alpha\beta}(\w)$ are the Born effective
  charges defined as the polarization along $\alpha$ induced by a displacement of the $s$-ion along $\beta$, and
  $C^0_{ss'\alpha\beta}(\w)$ is the interatomic
  force-constants matrix. (The superscript 0 indicates \textit{short-circuit} electrical boundary conditions, as detailed in Sec.~\ref{sec:extD}.)
  All three ingredients are 
    %All three quantities are 
    routinely computed via
DFPT~\cite{gonze_dynamical_1997,baroni_phonons_2001} in the static
limit, which is appropriate for non-magnetic insulators.
Generalizing this formula to metals requires a dynamical approach,
with a careful treatment of the low-frequency limit~\cite{cappelluti_charged-phonon_2012,binci_first-principles_2021,
    dreyer_nonadiabatic_2022,marchese_born_2024,wang_dynamical_2022}.
 % was first derived
 % analytically in the static limit for polar insulators~\cite{born_dynamical_1954}
 % and has recently been generalized to finite
 % frequency,
Here we build on these recent developments and calculate the aforementioned response functions
while retaining their full frequency dependence
%in this work we retain their full frequency dependence and evaluate them 
within TD-DFPT~\cite{gross_local_1985}.

  \subsection{Ehrenfest action functional}\label{sec:action}

We now present our unified variational description of plasmons and phonons, where 
  the lattice and electromagnetic degrees of freedom are treated on an equal footing.
  First, we extend the electromagnetic action of 
  Ref.~\cite{bertsch_real-space_2000} to include the lattice degrees of 
  freedom, arriving at the extended Ehrenfest 
  action~\cite{kwong_coupling_1987,todorov_time-dependent_2001}
  \begin{align}
      S({\bf A},\{{\bf R}_{s}\}) = \int_0^{T} dt \Bigg[ 
      &E(\psi,{\bf A},{\bf R}_{s}) 
      - i\hbar\sum_n f_n\braket{\psi_n}{\dot\psi_n} \nonumber\\
      &- \sum_s \frac{m_s}{2}\lvert\dot{{\bf R}}_s\rvert^2  
      - \frac{\Omega}{8\pi}\lvert\dot{\bf A}\rvert^2
      \Bigg].
      \label{eq:action}
  \end{align}
  Here ${\bf A}$ is the spatially uniform vector potential encoding the 
  macroscopic electric field in the velocity gauge and Gaussian units, ${\bm \E} = -\dot{\bf A}$, $\psi$ denotes the electronic wave functions and $\{{\bf R}_s\}$ are the 
  classical nuclear positions. 
  The last term accounts for the Maxwell energy of the free fields~\cite{stengel_electric_2009,romaniello_time-dependent_2005}, and plays the role of the classical kinetic energy that is associated with the electromagnetic degrees of freedom.
%  electromagnetic energy of the macroscopic field or equivalently macroscopic currents, which is not contained 
 (This term vanishes identically in standard ground-state calculations of the periodic crystal under 
 short-circuit EBC.)
%  in the density-functional energy $E$ as routinely computed within "short-circuits" boundary conditions.
   Equation~\eqref{eq:action} implies a velocity-gauge formulation of the coupling to macroscopic electric fields and provides a natural generalization to finite frequency of the static Lagrangian in Ref.~\cite{stengel_electric_2009}, extending its application to metallic systems.
   
  In the Kohn--Sham framework, the electronic energy 
  \begin{align}
      E_{\rm KS}(\psi,{\bf A},\{{\bf R}_{s}\}) &= E_{\rm Ew}(\{{\bf R}_{s}\}) + E_{\rm Hxc}(\psi) + \nonumber\\
      &+ E_{\rm kin}(\psi,{\bf A}) + E_{\rm ps}(\psi,{\bf A},\{{\bf R}_s\}) ,
      \label{eq:energy}
  \end{align}
  is written as the sum of an Ewald $E_{\rm Ew}$, a Hartree-exchange-correlation $ E_{\rm Hxc}$, a kinetic $E_{\rm kin}(\psi,{\bf A})$ and a pseudopotential $E_{\rm ps}(\psi,{\bf A},\{{\bf R}_s\})$ contribution~\cite{martin_electronic_2020}.
  The vector potential ${\bf A}$ enters through minimal substitution in both the kinetic 
  energy~\cite{mahan_many-particle_2000},
  \begin{align}
      E_{\rm kin}(\psi,{\bf A}) = \frac{1}{2m_e}\int [d{\bf k}]\sum_l f_{{\bf k}l}
      \ev{\lvert{\bf p} + e{\bf A}\rvert^2}{\psi_{{\bf k}l}},
  \end{align}
  where ${\bf p}$ and $m_e$ are the electronic momentum and mass, 
  and the pseudopotential energy,
  \begin{align}
      E_{\rm ps}(\psi,{\bf A},\{{\bf R}_s\}) = \int [d{\bf k}]\sum_l f_{{\bf k}l}
      \ev{V^{\rm ps}_{\tilde{\bf k} }(\{{\bf R}_s\})
      }{\psi_{{\bf k}l}},
  \end{align}
  where $V^{\rm ps}_{\bf k}$ is the cell-periodic pseudopotential operator,
  % {\color{blue}\begin{align}
  %   \hat{V}^{\rm ps}_{\bf k}     & =                 
  %   \int d{\bf r} d{\bf r}' 
  %   \sum_{sL} e^{-i(m_e \dot{\bf R}^{L}_{s}/\hbar)\cdot({\bf r}'-{\bf r})} e^{i {\bf k}\cdot({\bf r}'-{\bf r})}  \nonumber                                                                        \\
  %   &  v^{\rm ps}_{s}({\bf r}-{\bf R}^{L}_{s},{\bf r}'-{\bf R}^{L}_{s})\op{\bf r}{\bf r'}.
  %   \label{eq:Vps_def}
  % \end{align}}
  \begin{align}
    \hat{V}^{\rm ps}_{\bf k}     & =                 
    \int d{\bf r} d{\bf r}' 
    V^{\rm ps}({\bf r}',{\bf r}) e^{i {\bf k}\cdot({\bf r}'-{\bf r})}\op{\bf r}{\bf r'}  \nonumber                                                                        \\
    V^{\rm ps}({\bf r}',{\bf r}) & = \sum_{sL}  v^{\rm ps}_{s}({\bf r}-{\bf R}^{L}_{s},{\bf r}'-{\bf R}^{L}_{s}).
    % V^{\rm ps}({\bf r}',{\bf r}) & = \sum_{sL} e^{-i(m_e \dot{\bf R}^{L}_{s}/\hbar)\cdot({\bf r}'-{\bf r})} v^{\rm ps}_{s}({\bf r}-{\bf R}^{L}_{s},{\bf r}'-{\bf R}^{L}_{s}).
    \label{eq:Vps_def}
  \end{align}
  that is the standard pseudopotential contribution sampled at a shifted momentum-space vector
  \begin{equation}
    \tilde{\bf k} = {\bf k} + \frac{e}{\hbar}{\bf A} -\frac{m_e}{\hbar} \dot{\bf R}_s.
  \end{equation}  
  Note the additional dependence of the momentum-space shift on the nuclear
    velocity~\cite{stengel_rototranslational_2026}; we shall come back to this point in Sec.~\ref{sec:travel_pp}. The choice of $({\bf A},\{{\bf R}_s\})$ 
  as free parameters corresponds to short-circuit BC at clamped ions; all 
  other physical regimes are accessed via Legendre transforms, as described 
  below.

  \subsection{Equations of motion
  %for plasmon-phonon dynamics
  }\label{sec:extD}
  We now restrict ourselves to parameters oscillating at a single frequency, 
  $\lambda(t) = \lambda^\w e^{-i\w t}$, with a small positive imaginary part 
  $\w = \Re \omega + i\eta$ to represent the adiabatic switching-on of the 
  driving field.
  % \st{and to account for the irreversible energy transfer to the 
  % electronic subsystem.} 
  In this regime, the TD-DFPT 
  framework~\cite{gross_local_1985,corso_density-functional_1996} provides 
  the frequency-dependent response functions of 
  Eqs.~\eqref{eq:chi_el}--\eqref{eq:chi_vib} as second derivatives of 
  the action at its stationary point. 
  These can be arranged into a Hessian matrix, 
  \begin{align}
      % \det\bigl[\mathbb{S}(\w)\bigr] = 0, \qquad
      \mathbb{S}_{ij}(\w) = \frac{\partial^2 S}{\partial\lambda_i^{-\w}
      \,\partial\lambda_j^{\w}},
      \label{eq:hessian_A}
  \end{align}
  whose blocks are
  \begin{subequations}
  \begin{align}
      \frac{\partial^2 S}{\partial A^{\w}_\alpha\,\partial A^{-\w}_\beta} &=
      -\frac{\w^2\Omega}{4\pi}\bigl[\delta_{\alpha\beta} 
      + 4\pi\chi^{\rm el}_{\alpha\beta}(\w)\bigr],
      \label{eq:hess_AA}\\
      \frac{\partial^2 S}{\partial R^{\w}_{s\alpha}\,\partial R^{-\w}_{s'\beta}} &=
      -\w^2 m_s\delta_{ss'}\delta_{\alpha\beta} + C^0_{ss'\alpha\beta}(\w),
      \label{eq:hess_RR}\\
      \frac{\partial^2 S}{\partial R^{\w}_{s\beta}\,\partial A^{-\w}_\alpha} &=
      i\w\,\frac{\partial P_\alpha(\w)}{\partial R^{\w}_{s\beta}}
      = i\w e\,Z_{s\alpha\beta}(\w),
      \label{eq:hess_RA}\\
      \frac{\partial^2 S}{\partial A^{\w}_\alpha\,\partial R^{-\w}_{s\beta}} &=
      -i\w e\,Z_{s\alpha\beta}(-\w),
      \label{eq:hess_AR}
  \end{align}
  \label{eq:hess_blocks}
  \end{subequations}
  where Eq.~\eqref{eq:hess_AR} follows from Eq.~\eqref{eq:hess_RA} by 
  the reality of the susceptibilities in real space. 
  Equations~\eqref{eq:hess_AA} and~\eqref{eq:hess_RR} recover the 
  well-known equations of motion for the electric field and the lattice, 
  governed by the clamped-ion dielectric 
  response~\cite{mahan_many-particle_2000,ambegaokar_electromagnetic_1960} 
  and the short-circuit force constants, respectively. The mixed 
  derivatives Eqs.~\eqref{eq:hess_RA}--\eqref{eq:hess_AR} encode the 
  Born effective charges, %---simultaneously 
  which are defined as the polarization induced by an ionic displacement at linear order in the amplitude. In the dynamical case, they are related to the ionic force induced by an electric field via a time-reversal operation; note that $Z(\w)=Z(-\w)$ in presence of time-reversal symmetry.

  Eqs.~\eqref{eq:hessian_A}--\eqref{eq:hess_blocks} describe short-circuit, 
  clamped-ion conditions, which do not correspond directly to a realistic
  physical situation. 
  By combining the strategy of Ref.~\cite{royo_dynamical_2026} with that of Ref.~\cite{stengel_electric_2009},
  we can move between different electrical boundary conditions (open circuit,
  short circuit) and regimes (clamped ions, relaxed ions) via partial or
  total Legendre transforms, which, by the duality properties of Legendre 
  pairs, reduce to simple linear-algebra operations on the Hessian and 
  require no additional first-principles calculations.
  Equation~\eqref{eq:chi_vib} can be regarded as an example of such a procedure. By partially Legendre-transforming our original functional to relaxed-ion conditions, the $3\times 3$ head of the matrix acquires an extra term, carrying the lattice-mediated contributions to the dielectric tensor.
The resulting relaxed-ion short-circuit regime is appropriate, e.g., for simulating infrared absorption spectra.

To capture LO phonons and plasmons, all we need to do at this point is to impose open-circuit EBC along the propagation
  direction, $\hat{\mathbf{q}}$, while retaining short-circuit EBC in the perpendicular plane.
  This choice is consistent with the extremely oblate sample geometry
  underlying Eq.~\eqref{eq:inv_diel}~\cite{resta_polarization_2006,
  stengel_electric_2009}.
  Under these conditions, the plasmon-phonon interplay can be conveniently described
  by means of a reduced $(3N+1)\times(3N+1)$ Hessian ($N$ number of atoms per cell),
  where the additional row and column are
  obtained from
  Eq.~\eqref{eq:hessian_A} by projecting the 
  electric components of all quantities along the longitudinal direction, $\hat{\bf q}$. 
  Hereafter we refer to such projections as $\e^{\rm el}(\w)$ and $Z_{s\beta}(\omega) = {\bf Z}_{s\beta}(\omega)\cdot \hat{\bf q}$ for the dielectric function and Born charges, respectively.
  Then, by expressing ${\bf A}$ in terms of the electric field,
  $\bm{\mathcal{E}}^\omega = i\omega{\bf A}^\omega$, one obtains
  \begin{align}
      \bar{\mathbb{S}}(\omega) = -
      \begin{bmatrix}
          \dfrac{\Omega}{4\pi}\e^{\rm el}_{}(\omega)
          & e\,Z_{s'\beta'}(\omega) \\[2ex]
          e\,Z_{s\beta}(-\omega)
          & \omega^2 M_{ss'} - C^0_{ss'\beta\beta'}(\omega)
      \end{bmatrix},
      \label{eq:hessian_E}
  \end{align}
  in close analogy with the zero-frequency functional of
  Ref.~\cite{baroni_phonons_2001}, Eq.~(89).

  A total Legendre transform over both conjugate pairs,
  $(\mathcal{E}_\parallel,\{{\bf R}_s\}) \to
  (\mathcal{E}^0_\parallel,\{{\bf f}_s\})$~\cite{royo_dynamical_2026},
  amounts to inverting the reduced Hessian.
  % The head of $\bar{\mathbb{S}}^{-1}(\omega)$ is the total dielectric tensor \beqref{eq:inv_diel}.
  % Its poles are the coupled plasmon-phonon resonance frequencies under
  % open-circuit, relaxed-ion conditions.
  The $3N+1$ resonances under the mixed EBC for the propagation direction $\hat{\bf q}$,
  including all lattice modes, the longitudinal plasmon and possible hybrids with intermediate character,
  are the poles of $\bar{\mathbb{S}}^{-1}(\omega)$,
  identified by the following condition, which generalizes \beqref{eq:diel=0}:
  \begin{align}
      \det[\bar{\mathbb{S}}(\omega) ] = 0.
      \label{eq:eom}
  \end{align}
  % yielding plasmon-phonon hybrid excitation and transverse phonons.
  %
  The block structure of $\bar{\mathbb{S}}^{-1}(\w)$, obtained by block-wise inversion 
  of Eq.~\eqref{eq:hessian_E}, directly reveals the physical content. 
  The $(1,1)$ block is  $-(4\pi/\Omega)\,\epsilon^{-1}(\w)$, where $\epsilon^{-1}(\w)$ is the inverse dielectric function
  of Eq.~\eqref{eq:inv_diel}, including the lattice contribution. 
  The $(2,2)$ block is the phonon propagator with the macroscopic field relaxed, 
  defined by the \textit{open-circuit} force-constant matrix
  \begin{align}
      C_{ss'\alpha\beta}(\w) = C^0_{ss'\alpha\beta}(\w) 
      + \frac{4\pi e^2}{\Omega}\,Z_{s\alpha}(-\w)\,(\e^{\rm el})^{-1}(\w)
      \,Z_{s'\beta}(\w),
      \label{eq:ifc_na}
  \end{align}
  which reduces to the Cochran--Cowley result~\cite{cochran_dielectric_1962} 
  in the static limit of insulators, $(\e^{\rm el,bound})^{-1}(i0^+) = \epsilon_\infty^{-1}$. 
  The off-diagonal blocks can be written as $-({4\pi e}/{\Omega})\,Z^L_{s\alpha}(\w)\,[\w^2 M_{ss'} - 
  C_{ss'\alpha\beta}(\w)]^{-1}$, where $Z^L_{s\alpha}(\w) = 
  (\e^{\rm el})^{-1}(\w)\,Z_{s\alpha}(\w)$ are the longitudinal Callen effective 
  charges~\cite{ghosez_dynamical_1998}.

% \begin{align}
%   \begin{bmatrix} a & b \\ c & d \end{bmatrix}^{-1}
%   &= \begin{bmatrix} h & -hbd^{-1} \\ -d^{-1}ch & (d-ca^{-1}b)^{-1} 
%     \end{bmatrix}
%   \\ \text{with } h &\equiv (a - bd^{-1}c)^{-1},
%   \label{eq:matrix_inversion}
% \end{align}

In many cases of practical interest (e.g., a cubic crystal) the longitudinal and transverse directions are decoupled by symmetry. 
Then, the theory developed so far is appropriate to capturing both the transverse and the longitudinal low-energy spectrum by applying Eqs.~\eqref{eq:inv_diel} and~\eqref{eq:diel_tot} separately to each sector.
In sufficiently low-symmetry situations, however, the longitudinal and transverse modes are electrically coupled (e.g., a transverse mode generally induces longitudinal electric fields, and a longitudinal mode may be associated with a transverse polarization).
To treat such cases, the transverse and longitudinal spectra must be obtained as part of the same linear-algebra problem. 
An effective way of doing so is to retain the two additional rows and columns of the Hessian that correspond to the transverse electrical components.
%
%By applying a partial Legendre transform to the
By imposing mixed EBC on the 
resulting $(3N+3)\times(3N+3)$ matrix (this consists in imposing \textit{open-circuit} BC along the longitudinal direction and \textit{short-circuit} BC
along the transverse ones), the physically correct 
feedback between the longitudinal and transverse sectors is automatically propagated to the additional elements of the Hessian.
In practice, this is achieved via a partial Legendre transform involving only the longitudinal-field and
phonon blocks, and requires only a marginal modification of the procedure described above.
%     \st{
%     As a by-product, we also obtain the transverse infrared spectrum by keeping the full
% $(3N+3)\times(3N+3)$ matrix $\mathbb{S}_{ij}(\w)$ and imposing mixed boundary conditions:
% . A partial inversion of only the longitudinal-field and
% phonon blocks then covers both isotropic and anisotropic systems.
% In the trivial isotropic case, the two components decouple: changing the EBC leaves
% the transverse component unchanged, and the longitudinal and transverse responses
% follow from \eqref{eq:inv_diel} and \eqref{eq:diel_tot}, respectively.
% In anisotropic systems, by contrast, a transverse perturbation induces a longitudinal
% response, so that an echo of the longitudinal resonances appears in the transverse spectrum.
%     }
  % This formalism is more general than the description of plasmon-phonon
  % coupling in cubic systems. Keeping the full $(3N+3)\times(3N+3)$ matrix
  % $\mathbb{S}_{ij}(\w)$ and partially inverting its longitudinal and phonon
  % blocks yields the complete relaxed-ion dielectric tensor under mixed boundary
  % conditions: \textit{open-circuit} along the longitudinal direction and
  % \textit{short-circuit} along the transverse ones.
  % The transverse components thus acquire the vibrational contribution of
  % Eqs.~\eqref{eq:diel_tot}--\eqref{eq:chi_vib}. Moreover, treating the three
  % components of the macroscopic field within a single framework allows one to
  % describe anisotropic systems, in which a transverse perturbation induces a
  % longitudinal response and vice versa.

  \subsection{Generalized acoustic sum rule}\label{sec:sumrule}

  The present framework enables a rigorous analysis of
  momentum conservation in the coupled plasmon-phonon system, encoded in
  the generalized acoustic sum rule~\cite{dreyer_nonadiabatic_2022}.
  In insulators, the acoustic sum rule follows from the invariance of the
  system under a uniform rigid translation: since such a displacement
  induces no macroscopic polarization, the Born effective charges must
  satisfy~\cite{born_dynamical_1954}
  \begin{align}
      \sum_s Z_{s\alpha\beta}(0) = 0.
      \label{eq:asr_ins}
  \end{align}
  This implies that acoustic modes carry zero oscillator strength and are
  therefore decoupled from the dielectric response.

  In conducting systems this argument breaks down. A uniform translation
  of the lattice no longer rigidly displaces the entire system, because
  the delocalized electronic cloud does not follow the ions instantaneously.
  The electronic inertia introduces a finite, frequency-dependent response
  of the charge density to the ionic motion, modifying the sum rule
  to~\cite{dreyer_nonadiabatic_2022}
  \begin{align}
      \sum_s Z_{s\alpha\beta}(\w) = \frac{m_e\Omega}{e^2}
      \left[\frac{D_{\alpha\beta}}{\pi}
      - \w^2\chi^{\rm el,bound}_{\alpha\beta}(\w)\right].
      \label{eq:sumrule}
  \end{align}
  In the static limit, the right-hand side reduces to a term proportional
  to the Drude weight, which by comparison with Eq.~\eqref{eq:drude_jellium}
  is identified as the effective number of free carriers,
  \begin{align}
      \sum_s Z_{s\alpha\beta}(i0^+) =
      \frac{m_e\Omega}{e^2}\frac{D_{\alpha\beta}}{\pi}
      \equiv n^*_{\alpha\beta}.
      \label{eq:sumrule_stat}
  \end{align}
  At finite frequency, the interband term $\chi^{\rm el,bound}(\w)$ accounts
  for the oscillatory response of the bound electrons. 
  We may interpret \beqref{eq:sumrule} as the number (absolute value) and type (sign) of unbound carriers $n^*_{\alpha\beta}(\w)$ (see \ref{fig:sumrule}). 
  
  Physically,
  Eq.~\eqref{eq:sumrule} enforces total momentum conservation: in a
  charge-neutral system, any momentum acquired by the electronic subsystem
  under an external electric field must be exactly compensated by an equal
  and opposite ionic momentum, provided both subsystems are treated on an
  equal footing.

  In the $n$-type doped case, ionized donors add a positive background charge,
compensated by the same number of conduction electrons. We model this by adding
$|\delta\rho_{\rm dop}|$ electrons per unit cell together with a uniform positive
jellium background, as is routinely done in DFT codes. The Born effective charges
keep the usual decomposition into nuclear and electronic
contributions~\cite{marchese_born_2024,dreyer_nonadiabatic_2022},
\begin{align}
    Z_{s\alpha\beta}(\w) = Z^{\rm ion}_s\delta_{\alpha\beta} + Z^{\rm el}_{s\alpha\beta}(\w),
    \label{eq:becs_def}
\end{align}
and we assign the background charge to the site that the donors substitute. We also neglect
the small mass difference between the two elements. 
%  In BaSnO$_3$, La doping substitutes for Ba, so $Z^{\rm ion}_{\rm Ba} = 10 - \delta\rho_{\rm dop}$.

  The change in carrier density modifies the electronic contribution by shifting the Fermi level, as
  manifested in the electronic part of the sum rule
  \begin{align}
      \sum_s Z^{\rm el}_{s\alpha\beta}(i0^+)
      = -\bigl(N^{\rm el} - n^*\bigr)\delta_{\alpha\beta}
      \equiv -n^{\rm bound}_{\alpha\beta},
      \label{eq:sumrule_el}
  \end{align}
  where the minus sign takes into account the negative electronic charge and $n^{\rm bound}$ is the effective number of bound (non-Drude) electrons,
  defined as the difference between the total electron count and the free
  carrier number $n^*$.

  \subsection{Pseudo-static approximation}\label{sec:approx}

  The full frequency dependence of the response functions entering
  $\mathbb{S}(\w)$ is in general costly to compute. The common practice for conducting systems is therefore to adopt a pseudo-static approximation,
  valid when the electronic interband gap greatly exceeds both the phonon
  and plasma frequencies. In this limit the frequency dependence of the
  interatomic force constants, the Born effective charges, and the interband
  dielectric response can all be neglected, while the analytic Drude
  dependence is retained to account for the metallic character of the system:
  \begin{subequations}
    \begin{align}
      {\epsilon}^{\rm el}_{\alpha\beta}(\w) & \approx
      \delta_{\alpha\beta} - \frac{4{D}_{\alpha\beta}}{\w^2} + 4\pi{\chi}^{\rm el,bound}_{\alpha\beta}(i0^+), \\
      C^0_{ss'\alpha\beta}(\w)              & \approx C^0_{ss'\alpha\beta}(i0^+),                          \\
      Z_{s\alpha\beta}(\w)                  & \approx Z_{s\alpha\beta}(i0^+).
    \end{align}
    \label{eq:stat_approx}
  \end{subequations}
This approximation reduces 
the coupled electron-phonon problem to a secular equation, in close
analogy to the approach of Ref.~\cite{lihm_plasmon-phonon_2024};
in this sense, at $\mathbf q\to0$ the latter can be regarded as a limiting case of our scheme.
Note, however, that Eq.~\eqref{eq:stat_approx} differs from Ref.~\cite{lihm_plasmon-phonon_2024}
in two crucial aspects. First, it accounts for possible deviations in the Born effective charges, 
bound-charge dielectric susceptibility and interatomic force constants compared to the undoped case; in the previous 
Section we have seen an example of how such deviations may be essential to comply with basic
conservation laws. Second, the plasmon-pole structure of the longitudinal charge fluctuations
emerges at $\mathbf q\to0$ from the Drude term of $\epsilon^{\rm el}(\w)$, rather than entering as a model assumption.
Last but not least, our methodological framework allows for a systematic test of this 
approximation against the exact TD-DFPT result at a modest computational cost, as 
we shall see in Sec.~\ref{sec:method}.

 % \st{This approach was extended in Ref.%~\cite{lihm_plasmon-phonon_2024} to
 %  include the frequency dependence of the carrier response, approximated by a
 %  plasmon-pole model without local-field effects to reduce the computational
 %  cost. Using the interpolation scheme discussed in Sec.%~\ref{sec:interp}, we
 %  instead include the full frequency dependence of all ingredients and the
 %  full \textit{ab initio} dielectric response, at a modest computational cost.}

  \subsection{Minimal model of plasmon-phonon interplay}\label{sec:model}

  The pseudo-static approximation in Eq.~\eqref{eq:stat_approx} provides the starting point 
  for a minimal model of the plasmon-phonon coupling, comprising one acoustic
  branch $(\omega_A=0)$, one optical branch $(\omega_{\rm TO})$, and the electronic dielectric response
  \begin{align}
      \mathbb{S}^{\rm mod}(\w) =
      \begin{bmatrix}
        \frac{\Omega}{4\pi} \big[\w^2  \epsilon^{\rm el,bound} - 4D \big] & i\w Q_A & i\w Q_{\rm TO} \\[2ex]
          % \dfrac{\Omega}{4\pi}\e^{\rm el}(\w) & Q_A(0) & Q_{\rm TO}(0) \\[2ex]
          -i\w Q_A & \w^2 & 0 \\[1ex]
          -i\w Q_{\rm TO} & 0 & \w^2 - \omega_{\rm TO}^2
      \end{bmatrix}.
      \label{eq:hessian_model}
  \end{align}
  Here  Eq.~\eqref{eq:hessian_A} is rewritten in the basis that diagonalizes the
   dynamical matrix   $C^0_{ss'\alpha\beta}(0)/ \sqrt{M_s M_{s'}}$,
  with orthonormal eigenvectors $\xi^\mu_{s\alpha}$.
  The Drude term is singled out of the dielectric tensor by introducing
  $\epsilon^{\rm el,bound}=1+4\pi\chi^{\rm el,bound}(\w) \big|_{\w=0}$.
  The oscillator strength $Q_\mu = e\,\xi^\mu_{s\beta}(\w)\,
  \hat{q}_\alpha\,Z_{s\alpha\beta}(\w)/\sqrt{m_s} \big|_{\w=0}$ measures the coupling of phonon mode
  $\mu$ to the electric field. For the acoustic branch it reduces to
  $Q_A = (e/\sqrt{m^{\rm ion}_{\rm tot}})\sum_s Z_s(i0^+)$, with
  $m^{\rm ion}_{\rm tot} = \sum_s m_s$.
  Applying \beqref{eq:eom} to this case yields 
  \begin{align}
    [(\omega_p^2 +\Delta_A -\w^2) (\omega_{\rm TO}^2 -\w^2) -\Delta_{\rm TO}\w^2 ]\w^2= 0
  \end{align}
  with the plasma frequency
  $
      \omega_p^2 = 4D/\epsilon^{\rm el,bound}
      % \label{eq:plasma_bare}
  $ and the screened oscillator strength $\Delta_{\mu}=\frac{4\pi}{\Omega} Q_\mu^2/\epsilon^{\rm el,bound}$.
  The oscillator strength of the acoustic mode enters then as an additive correction to the squared plasma frequency  
  \begin{align}
      \bigl(\omega_p^{\rm tot}\bigr)^2 =
      \omega_p^2 + \frac{4\pi}{\Omega}
      \frac{Q_A^2}{\epsilon^{\rm el,bound}}.
      \label{eq:plasma_tot}
  \end{align}

\begin{figure}[h]
    \centering
    \includegraphics[width=0.95\linewidth]{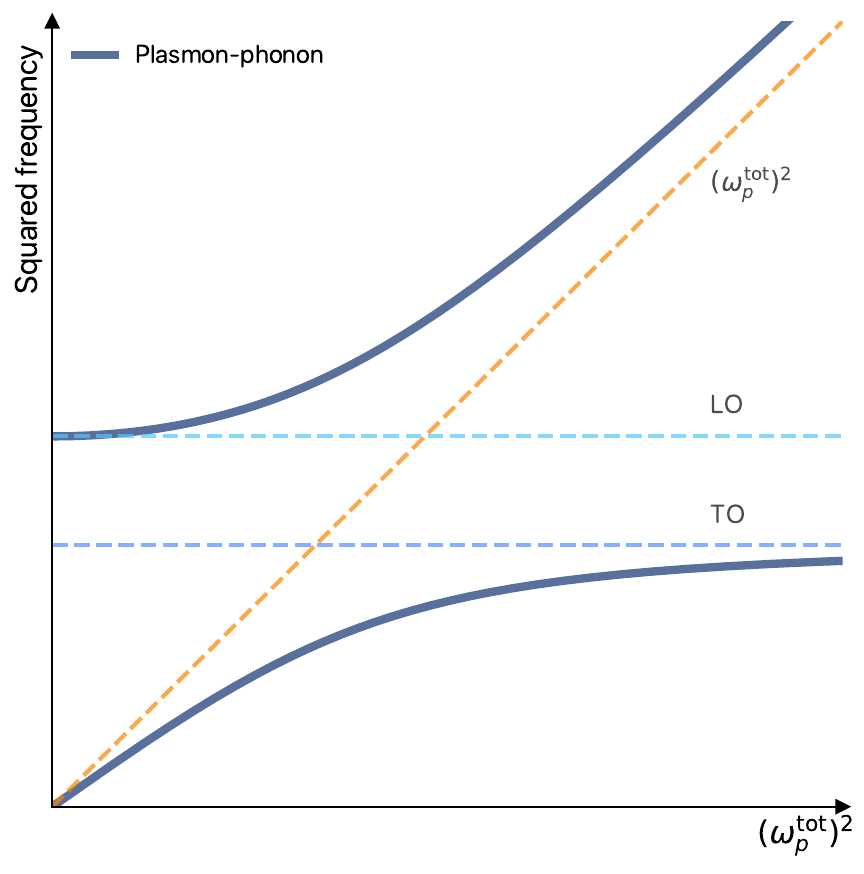}~%
\caption{Plasmon-phonon modes of the minimal model,
  Eq.~\eqref{eq:model_roots}, as a function of doping, parametrized by the
  total plasma frequency $\omega_p^{\rm tot}$ of
  Eq.~\eqref{eq:plasma_tot}; frequencies are shown squared. % and $\omega_{\rm TO}=1$ $\Delta_{\rm TO}=1$. 
  The two  branches (solid) display an avoided crossing: the lower one evolves from
  the plasmon at low doping into the TO phonon, the upper one from the LO
  phonon into the plasmon. No longitudinal mode exists within the
  Reststrahlen band between the static TO and LO frequencies (horizontal
  dashed lines). The diagonal dashed line is the plasma frequency
  renormalized by the acoustic branch but not coupled to the optical
  phonon.}
    \label{fig:model}
  \end{figure}

  % Resolving for the plasma frequency shows that
  % the bare plasma frequency
  % \begin{align}
  %     \omega_p^2 = \frac{4D}{1 + 4\pi\chi^{\rm el,bound}(0)}
  %     \label{eq:plasma_bare}
  % \end{align}
  % acquires an additive correction from the acoustic branch,
  % \begin{align}
  %     \bigl(\omega_p^{\rm tot}\bigr)^2 =
  %     \omega_p^2 + \frac{4\pi}{\Omega}
  %     \frac{Q_A(0)^2}{1 + 4\pi\chi^{\rm el,bound}(0)},
  %     \label{eq:plasma_tot}
  % \end{align}
The model can then be solved analytically, giving
  \begin{align}
    \w^2_{\pm} = \frac{B \pm \sqrt{B^2-4\bigl(\omega_p^{\rm tot}\bigr)^2\omega_{\rm TO}^2}}{2}
    \label{eq:model_roots}
  \end{align}
  with $B=\bigl(\omega_p^{\rm tot}\bigr)^2+\omega_{\rm TO}^2+\Delta_{\rm TO}$.
  Assuming that the Born effective charges and the interband dielectric tensor are weakly dependent on the doping, we can study the qualitative behavior of the two solutions in \beqref{eq:model_roots} as a function of the plasma frequency $\omega_p$ (see \ref{fig:model}).
   At very small carrier densities (limit of $\omega_p \to 0$), the doped semiconductor behaves exactly like the undoped crystal. For nonzero but small plasma frequency ($\omega_p \ll \omega_{\rm TO}$), the plasmon ``sees'' the additional dielectric screening by the optical phonon (the latter acts as an additional bound charge that responds dielectrically to the field), so its frequency is $\omega\simeq \sqrt{4D/\epsilon_{\rm st}}$ with 
   \begin{align}
    \epsilon_{\rm st} = \epsilon^{\rm el,bound} + \frac{4\pi}{\Omega}\frac{Q_{\rm TO}^2}{\omega_{\rm TO}^2}.
   \end{align}
   In the opposite limit of large plasma frequency, we have a \textit{reverse} LO-TO \textit{splitting regime}, where the LO phonon becomes softer than the TO. The shift in the frequency squared amounts to
  \begin{align}
    \bar{\Delta}(\w) = \omega_{\rm LO}^2(\w) - \omega_{\rm TO}^2 = \frac{4\pi}{\Omega}\frac{Q_{\rm TO}^2}{\epsilon^{\rm el,bound}-4D/\w^2}
  \end{align}
   i.e., the same formula as in the insulating case, with the important difference that, in this regime, the plasmon makes the denominator negative, resulting in $\bar{\Delta}(\w)<0$.
   Thus, the LO phonon reaches its asymptotic TO value from below~\cite{varga_coupling_1965}.

  \subsection{Plasmon renormalization by lattice inertia}\label{sec:pla_renorm}

  Momentum conservation, expressed by the generalized acoustic sum rule
  Eq.~\eqref{eq:sumrule}, constrains the relative motion of the electronic
  and ionic subsystems in a transparent way. Under an external electric
  field, the negative electronic charges and the positive ions are
  displaced in opposite directions; however, since they carry opposite
  charges, the associated currents flow in the same direction and their
  contributions to the total conductivity add rather than cancel. The net
  effect is an enhancement of the Drude weight --- and an upward
  renormalization of the plasma frequency --- when the ionic positions are
  allowed to relax. This can already be seen in the model introduced 
  in the previous section.

  Using the acoustic sum rule, Eq.~\eqref{eq:sumrule_stat}, the total plasma frequency of \beqref{eq:plasma_tot}
  simplifies to
  \begin{align}
      \bigl(\omega_p^{\rm tot}\bigr)^2 = 
      \omega_p^2\left(1 + \frac{n^* m_e}{m^{\rm ion}_{\rm tot}}\right),
      \label{eq:plasma_vib}
  \end{align}
  which generalizes, beyond free-electron-like metals to a crystal with electronic band structure, the 
  Bohm--Staver relation relating the ``sound'' velocity of acoustic modes $v_{\rm ac}$ and the Fermi velocity $v_{F}$~\cite{bohm_application_1951,ashcroft_solid_1976},
  $v^2_{\rm ac} = (N^{\rm el}m_e/3m^{\rm ion})v_F^2$. The key
  difference is that the total electron count $N^{\rm el}$ of the free-electron result is replaced by the effective free-carrier number $n^*$,
  which excludes electrons localized in bound states.
  The underlying physical picture follows directly from momentum
  conservation: fixing the center of mass of the charge-neutral system
  constrains the response to involve only relative motion, so that the
  relevant inertia is the reduced mass of the electron-ion system. The
  total Drude weight of the coupled system, generalizing the free-electron
  result of \beqref{eq:drude_jellium} to include ionic inertia, is
  \begin{align}
      D^{\rm tot} = \frac{\pi e^2}{\Omega}(n^*)^2
      \left[\frac{1}{n^* m_e} + \frac{1}{m^{\rm ion}_{\rm tot}}\right]
      = \frac{\pi e^2}{\Omega}\frac{(n^*)^2}{\mu},
      \label{eq:drude_tot}
  \end{align}
  where $\mu = n^*m_e\,m^{\rm ion}_{\rm tot}/(n^*m_e + m^{\rm ion}_{\rm tot})$
  is the reduced mass of the coupled system. 

  In the finite-frequency case, the effective plasma frequency is
  renormalized by the cumulative oscillator strength of all phonon modes
  with frequencies well below $\omega_p$,
  \begin{align}
      \bigl(\omega_p^{\rm eff}\bigr)^2 =
      \omega_p^2 + \frac{4\pi}{\Omega}\sum_{\omega_\mu \ll \omega_p}
      \frac{Q_\mu^2}{\epsilon^{\rm el,bound}},
      \label{eq:plasma_eff}
  \end{align}
  while modes with frequencies comparable to $\omega_p$ contribute in more
  complex ways, giving rise to hybrid excitations rather than a simple
  additive shift. 
  % Neglecting these corrections leads to a systematic
  % underestimation of the electronic Drude weight when it is extracted from
  % the plasma frequency in experimental reflectivity spectra
  % (see \ref{fig:bso_R}).

  Going beyond the pseudo-static approximation, the bound electrons
  actively participate in the electron-phonon coupling, enhancing the
  effective lattice inertia. This is evidenced by the translational sum
  rule for the force-constants matrix~\cite{stengel_rototranslational_2026},
  \begin{align}
      \sum_{ss'} C^0_{ss'}(\w) =
      \sum_{ss'} C^0_{ss'}(0) - \w^2 m_e n^{\rm bound} + \mathcal{O}(\w^3),
      \label{eq:ifc_sumrule}
  \end{align}
  which shows that the low-frequency correction to the adiabatic ionic masses~\cite{scherrer_mass_2017} is controlled by $n^{\rm bound}$, the number of bound
  electrons defined in Eq.~\eqref{eq:sumrule_el}. 

  \section{Computational method}\label{sec:method}

  \subsection{Frequency-dependent DFPT}\label{sec:numerical}

  The extended dynamical matrix $\mathbb{S}(\w)$ introduced in Sec.~\ref{sec:extD} 
  requires as input the frequency-dependent Born effective charges $Z_{s\alpha\beta}(\w)$, 
  interatomic force constants $C^0_{ss'\alpha\beta}(\w)$, and electronic dielectric response 
  $\chi^{\rm el}_{\alpha\beta}(\w)$. While these quantities are routinely computed in the 
  static limit via DFPT~\cite{baroni_phonons_2001}, 
  retaining their full frequency dependence is essential for treating 
  plasmon-phonon coupling. We achieve this by extending the time-dependent DFPT framework 
  of Ref.~\cite{binci_first-principles_2021} to include the complete frequency 
  dependence of the self-consistent potential, thereby generalizing the ensemble DFPT 
  approach~\cite{zabalo_ensemble_2024} to finite frequency. More technical details and the finite momentum generalization are described in \cite{paper2}.

  The general susceptibility 
  relating observables $A$ and $B$ takes the form
  \begin{align}
    \chi_{AB}(\w) 
    &= \Tr[H^{AB}_{\rm bare}\; \rho]
    + \Tr[H^{A}_{\rm bare}\; \rho^{B}(\w)].
    %  \nonumber \\
    % &= \frac{s_d}{N_{\mathbf{k}}}\sum_{\mathbf{k}l}
    % \Bigl[
    %   \mel{\psi_{\mathbf{k}l}}{H^{A}_{\rm bare}}{u^{B}_{\mathbf{k}l}(\w)} \nonumber \\
    %   &\qquad + \mel{u^{B}_{\mathbf{k}l}(\w)}{H^{A}_{\rm bare}}{\psi_{\mathbf{k}l}}
    %   \Bigr],
    \label{eq:chi_AB}
  \end{align}
  expressed as the traces of the \textit{bare}---without electronic screening---perturbation operators $H^{A}_{\rm bare}$ and $H^{AB}_{\rm bare}$
  times the equilibrium 
  \begin{align}
    \rho &= 
    \int [d{\bf k}]  \sum_{l}   f_{\mathbf{k}l}  \op{\mathbf{k}l}
  \end{align}
  and the perturbed 
  density operator 
  \begin{align}
  \rho^{B}(\w) &= \int [d{\bf k}]
    \sum_l %_{l=1}^M 
    f_{\mathbf{k}l}\left(   
    \ket{\psi_{\mathbf{k}l}^B(\w)} \bra{\mathbf{k}l}+
    \op{\mathbf{k}l}{\psi_{\mathbf{k}l}^B(-\w^*)}
    \right) 
    \label{eq:ind_density}
  \end{align} 
  where $\ket{\mathbf{k}l}$  denotes  unperturbed Bloch states 
  with momentum ${\bf k}$, band energies 
  $\e_{\mathbf{k}l}$ and occupations $f_{\mathbf{k}l}$.
  Calculating $\rho^{B}(\w)$ is the key computational challenge.
  To this end, we partition the Hilbert space into an \textit{active} 
  subspace of $M$ bands near the Fermi level, where transitions contribute 
  most strongly to the response, and a \textit{passive} complement containing 
  high-energy transitions. This decomposition is defined by the projector
  \begin{align}
  \mathcal{P}_{\mathbf{k}} 
  = \sum_{l=1}^M 
    % \op{l}{l} 
  \op{\mathbf{k}l}
    \label{eq:proj}
  \end{align}
  and allows us to separate two contributions to the first-order wave functions
  \begin{align}
    \ket{\psi^{B}_{\mathbf{k}l}(\w)} =
    \ket{\psi^{B}_{\mathbf{k}l}(\w)}_{\rm a} +
    \ket{\psi^{B}_{\mathbf{k}l}(\w)}_{\rm p}
  \end{align}
  and to the induced density operators
  \begin{align}
    % \rho(\w) &= 
    % \int [d{\bf k}]  \sum_{l}   f_{\mathbf{k}l}  \op{\mathbf{k}l}
    % \\
    \rho^{B}(\w) &= %\rho^{B}(\w) \mathcal{P}_{\mathbf{k}} + \rho^{B}(\w)(1- \mathcal{P}_{\mathbf{k}})  = 
    \rho^{B}_a(\w) + \rho^{B}_p(\w).
  \end{align}
  The advantage of this splitting is that the two terms can be computed with different strategies.
  The passive-subspace contribution is obtained by solving the Sternheimer
  equation 
  \begin{align}
    \bigl(H_{\mathbf{k}} + \alpha\mathcal{P}_{\mathbf{k}}
    - \e_{\mathbf{k}l} - \hbar \w\bigr)
    \ket{\psi_{{\bf k}l}^{B}(\w)}_{\rm p} 
    % \ket{u^{B}_{\mathbf{k}l}(\w)}_{\rm p} 
      = \nonumber\\
      -(1-\mathcal{P}_{\mathbf{k}})\,
    H^{B}_{\rm scf}(\w)\,\ket{{\bf k}l}, %\ket{\psi_{\mathbf{k}l}},
    \label{eq:stern}
  \end{align}
  where 
  the level-shift term $\alpha\mathcal{P}_{\mathbf{k}}$ removes singularities associated with 
  the active-subspace states, analogously to the static case~\cite{de_gironcoli_lattice_1995}. 
  The screened perturbation Hamiltonian 
  $H^{B}_{\rm scf}(\w)$ acquires a frequency dependence from self-consistency via the exchange-correlation 
  and Hartree kernels.
  We solve Eq.~\eqref{eq:stern} iteratively using a biconjugate-gradient algorithm, 
  which preserves the complex-valued solution at finite frequency.

  The active-subspace contribution is evaluated explicitly via a Kubo-like 
  sum over bands, 
  \begin{align}
    \rho^{B}_a(\w) &= \int [d{\bf k}] 
        \sum_{l,m=1}^M F_{{\bf k} ml}(\w)
        \ket{\mathbf{k}l}
        \bigl({H^B_{\rm scf}(\w)}\bigr)_{\mathbf{k}lm}
        \bra{\mathbf{k}m}
    \label{eq:rho_active}
  \end{align}
  repacked into the active-space first-order wave functions, as routinely done for the static case~\cite{de_gironcoli_lattice_1995}, 
  \begin{align}
    \ket{\psi^{B}_{\mathbf{k}l}(\w)}_{\rm a} = f_{\mathbf{k}l}^{-1}
    \sum_{m=1}^{M}\theta_{\mathbf{k}lm}\,F_{\mathbf{k}lm}(\w)\,
      \bigl({H^B_{\rm scf}(\w)}\bigr)_{\mathbf{k}ml}
    % \mel{\mathbf{k}m}{H^{B}_{\rm scf}(\w)}{\mathbf{k}l}
    \ket{\mathbf{k}m},
  \end{align}
  where ${(O)_{\mathbf{k}lm} = \mel{\mathbf{k}l}{O}{\mathbf{k}m}}$, the Heaviside function $\theta_{\mathbf{k}lm} = \theta(\e_{\mathbf{k}l} -
    \e_{\mathbf{k}m})$ enforces band ordering, and 
  \begin{align}
    F_{\mathbf{k}lm}(\w) =  
        \dfrac{f_{\mathbf{k}l} - f_{\mathbf{k}m}}
        {\e_{\mathbf{k}l} - \e_{\mathbf{k}m}
          + \hbar \w}
  \end{align}
  is the frequency-dependent electron-hole propagator, which 
  has only interband contributions, in contrast to the adiabatic electron-hole propagator
  \begin{align}
      \bar{F}_{\mathbf{k}lm} = \begin{cases}
          \dfrac{1}{2}\bigl(f'_{\mathbf{k}l} + f'_{\mathbf{k}m}\bigr),
          & \e_{\mathbf{k}l} = \e_{\mathbf{k}m}, \\[6pt]
          \dfrac{f_{\mathbf{k}l} - f_{\mathbf{k}m}}
          {\e_{\mathbf{k}l} - \e_{\mathbf{k}m}},
          & \e_{\mathbf{k}l} \neq \e_{\mathbf{k}m},
      \end{cases}
  \end{align}
  which includes the intraband ($\e_{\mathbf{k}l} =
  \e_{\mathbf{k}m}$) contribution through the occupation
  derivative $f'_{\mathbf{k}l}$.

  % The corresponding active and passive contributions to the induced density operators are
  % \begin{align}
  %   \rho^{B}_a(\w) &= \int [d{\bf k}] 
  %     \sum_{ml=1}^M \frac{f_{\mathbf{k}l}-f_{\mathbf{k}m}  }   
  %     {\e_{\mathbf{k}l}-\e_{\mathbf{k}m} + \omega}
  %     \ket{\mathbf{k}l}
  %     \bigl({H^B_{\rm scf}(\w)}\bigr)_{\mathbf{k}lm}
  %     \bra{\mathbf{k}m} \nonumber \\
  %   \rho^{B}_p(\w) &= \int [d{\bf k}]
  %   \sum_{l=1}^M 
  %    f_{\mathbf{k}l}\left(   
  %   \ket{\psi_{\mathbf{k}l}^B(\w)}_{\rm p} \bra{\mathbf{k}l}+
  %   \op{\mathbf{k}l}{\psi_{\mathbf{k}l}^B(-\w^*)}_{\rm p}
  %   \right) 
  %   \label{eq:ind_density_sep}
  % \end{align}

  % The complex frequency 
  % $\w = \omega + i\eta$ with $\eta > 0$ ensures causality and represents the 
  % adiabatic switching-on of the perturbation.

  \subsection{Response to electric field}\label{sec:efield}

  The electronic dielectric response in the velocity gauge requires careful
  treatment of the interplay between diamagnetic and paramagnetic
  contributions. We derive it here from the second derivative of the action
  Eq.~\eqref{eq:action} with respect to the vector potential, following an
  approach analogous to the finite-wavevector analysis of
  Ref.~\cite{ambegaokar_electromagnetic_1960}.

  From Eq.~\eqref{eq:hess_AA}, the short-circuit response to a vector
  potential involves both a diamagnetic term, arising from the quadratic
  dependence of $E_{\rm kin}$ on ${\bf A}$, and a paramagnetic term from
  the current-field coupling:
  \begin{align}
      \frac{\partial^2 S}{\partial A^{-\w}_\alpha\,\partial A^{\w}_\beta}
      &= -\frac{\Omega\w^2}{4\pi}\,\delta_{\alpha\beta}
      \nonumber\\
      &+ \Tr\bigl[H^{A_\alpha}_{\rm bare}\,\rho^{A_\beta}(\w)\bigr]
      + \Tr\bigl[H^{A_\alpha,A_\beta}_{\rm bare}\,\rho\bigr].
      \label{eq:hess_AA_expl}
  \end{align}
  A key simplification emerges from the $f$-sum rule, which enforces a partial 
  cancellation between these terms. Decomposing the
  response to a dynamical vector potential into a static and a frequency-dependent part via the relation between electric field $\E_\beta$ and vector potential $A_\beta$
  \begin{align}
      H^{A_\beta}_{\rm scf}(\w) &=  H^{A_\beta}_{\rm bare} + i\w\,H^{\E_\beta}_{\rm hxc}(\w)
      \label{eq:H_A_to_E}
      \\
      \rho^{A_\beta}(\w) &= \rho^{A_\beta}(0) + i\w\,\rho^{\E_\beta}(\w),
      \label{eq:rho_A_to_E}
  \end{align}
  the static contributions in Eq.~\eqref{eq:hess_AA_expl} can be combined
  and recast, through the minimal coupling ${\bf k}\to{\bf k}+
  (e/\hbar){\bf A}$, as a $\mathbf{k}$-space total derivative that
  integrates to zero over the Brillouin
  zone~\cite{ashcroft_solid_1976,melrose_electromagnetic_1991,
  ambegaokar_electromagnetic_1960}. The remaining terms reduce to
  \begin{align}
      \frac{4\pi}{\Omega}\frac{\partial^2 S}{\partial A^{-\w}_\alpha\,
      \partial A^{\w}_\beta} =
      -\w^2\delta_{\alpha\beta}
      - 4\pi i\w\,\sigma^{\rm el}_{\alpha\beta}(\w),
      \label{eq:hess_AA_cond}
  \end{align}
  where the full optical conductivity is defined as
  \begin{align}
      \sigma^{\rm el}_{\alpha\beta}(\w) \equiv
      -\frac{1}{\Omega}
      \Tr\bigl[H^{A_\alpha}_{\rm bare}\,\rho^{\E_\beta}(\w)\bigr].
  \end{align}

  For the numerical evaluation of $\sigma^{\rm el}_{\alpha\beta}(\w)$, 
  we adopt the same dual-subspace partition introduced in Sec.~\ref{sec:numerical}.
  Via a straightforward application of Eq.~\eqref{eq:rho_A_to_E} to the formalism described
  earlier, we obtain the induced density matrix at frequency $\w$ as $\rho^{\Ea}(\w) = \rho^{\Ea}_{\rm a}(\w) + \rho^{\Ea}_{\rm p}(\w)$, where the active-space (a) and passive-space (p) contributions are given by
  \begin{widetext}
  \begin{align}
      \rho^{\Ea}_{\rm a}(\w) =
      &\int[d{\bf k}]\sum_{l,m=1}^{M} \left(
      \frac{ie \bar{F}_{\mathbf{k}lm}}
      {\e_{\mathbf{k}m} - \e_{\mathbf{k}l} + \hbar\w}
      \ket{\mathbf{k}l}\bigl(H^{k_\alpha}_{\rm bare}\bigr)_{\mathbf{k}lm} \bra{\mathbf{k}m} 
      +
      % \int[d{\bf k}]\sum_{l,m=1}^{M}
      % \frac{f_{\mathbf{k}l} - f_{\mathbf{k}m}}
      % {\e_{\mathbf{k}l} - \e_{\mathbf{k}m} + \hbar\w}
      {F}_{\mathbf{k}ml}(\w)
      \ket{\mathbf{k}l}\bigl(H^{\Ea}_{\rm hxc}\bigr)_{\mathbf{k}lm}\bra{\mathbf{k}m} \right)
      \nonumber\\
      \rho^{\Ea}_{\rm p}(\w) = & \int[d{\bf k}]\sum_l f_{\mathbf{k}l}
      \Bigl(
      \ket{\psi^{\Ea}_{\mathbf{k}l}(\w)}_p \bra{\mathbf{k}l}
      + \ket{\mathbf{k}l}\bra{\psi^{\Ea}_{\mathbf{k}l}(-\w^*)}_p
      \Bigr),
      \label{eq:rhoE_full}
  \end{align}
  \end{widetext}

  In the ``p'' sector, $ \ket{\psi^{\Ea}_{\mathbf{k}l}(\w)}_p$ is the passive-subspace contribution to the first-order electric-field wave functions, which are defined by the Sternheimer equation
  \begin{align}
      \bigl(H_{\mathbf{k}} + \alpha\mathcal{P}_{\mathbf{k}}
      - \e_{\mathbf{k}l} - \hbar\w\bigr)
      \ket{\psi^{\Ea}_{\mathbf{k}l}(\w)}_{\rm p}
      &= \nonumber\\
      -(1-\mathcal{P}_{\mathbf{k}})
      \Bigl[ie\ket{\psi^{k_\alpha}_{\mathbf{k}l}}_p
      + &H^{\Ea}_{\rm hxc}(\w)\ket{\mathbf{k}l}\Bigr].
      \label{eq:stern_E}
  \end{align}
  As in the insulating case, the role of the external perturbation is played by 
  the (passive-subspace part of the) static ($\w=0$) momentum-derivative
  wave functions, $\ket{i\psi^{k_\alpha}_{\mathbf{k}l}}_{\rm p}$.
  The latter, in turn, are defined by
  %
  %\begin{align}
  %    \ket{i\psi^{k_\alpha}_{\mathbf{k}l}} =
  %    \ket{i\psi^{k_\alpha}_{\mathbf{k}l}}_{\rm p} +
  %    \ket{i\psi^{k_\alpha}_{\mathbf{k}l}}_{\rm a},
  %\end{align}
  %the passive contribution satisfies the static 
  the following Sternheimer equation
  \begin{align}
      \bigl(H_{\mathbf{k}} + \alpha\mathcal{P}_{\mathbf{k}}
      - \e_{\mathbf{k}l}\bigr)
      \ket{\psi^{k_\alpha}_{\mathbf{k}l}}_{\rm p} =
      -(1-\mathcal{P}_{\mathbf{k}})\,
      H^{k_\alpha}_{\rm bare}\ket{\mathbf{k}l}.
  \end{align}

  Regarding the active-space Kubo-like part of \beqref{eq:rhoE_full}, note that
  it contains an intraband contribution, which has been introduced when 
  passing from $\rho^{A_\beta}$ to $\rho^{\E_\beta}$ in
  Eq.~\eqref{eq:rho_A_to_E},
  %introduces an intraband contribution to the
  %electric-field density matrix that is absent in
  %$\rho^{A_\beta}(\w)$,
  \begin{align}
      \rho^{\E_\beta}_{\rm intra}(\w) = \frac{-e}{i\hbar\w}
      \int[d{\bf k}]\sum_l
      f'_{\mathbf{k}l}\,H^{k_\beta}_{\rm bare}\op{\mathbf{k}l},
  \end{align}
  with $f'_{\mathbf{k}l} = \partial f_{\mathbf{k}l}/
  \partial\e_{\mathbf{k}l}$. 
  This term diverges in the low-frequency limit, and 
  therefore it is natural to separate it from the remaining 
  interband contributions to $\rho^\Ea_{\rm a}(\w)$.
  In time-reversal-symmetric systems, $\rho^{\E_\beta}_{\rm intra}(\w)$ carries a
current but no charge: $\partial_{k_\beta}\e_{\mathbf{k}l}$ is odd in $\mathbf{k}$,
whereas $|\psi_{\mathbf{k}l}(\mathbf{r})|^2$ is even. It therefore does not contribute
to $H^{\E}_{\rm Hxc}$, and only the interband terms enter the self-consistent cycle. 
  The interband part, in turn, is conveniently written in a 
  form that closely resembles the active-space contribution of \beqref{eq:rho_active},
  \begin{equation}
  \begin{split}
    \rho^{\Ea}_{\rm a,inter}(\w) =&\\ \int [d{\bf k}]
      \sum_{\substack{l,m=1\\l\ne m}}^M&  
      F_{\mathbf{k}ml}(\w)
      % \frac{f_{\mathbf{k}l}-f_{\mathbf{k}m}  }   
      % {\e_{\mathbf{k}l}-\e_{\mathbf{k}m} + \hbar\w} \\
     \ket{\mathbf{k}l}
      \bigl({H^\Ea_{\rm scf}(\w)}\bigr)_{\mathbf{k}lm}
      \bra{\mathbf{k}m}.
    \end{split}
  \end{equation}
  The double summation over active states now excludes
  $m=l$.
  The perturbation consists, as usual, of a bare 
  contribution and a self-consistent (SCF) potential contribution, 
  $H^\Ea_{\rm scf}(\w) = H^\Ea_{\rm bare} + H_{\rm Hxc}^\Ea(\w)$, 
  with 
  \begin{align}
      \bigl(H^{\Ea}_{\rm bare}\bigr)_{\mathbf{k}lm}=
      \frac{ie\,\bigl(H^{k_\alpha}_{\rm bare}\bigr)_{\mathbf{k}lm}}
      {\e_{\mathbf{k}m} - \e_{\mathbf{k}l}},
      \label{eq:A_to_E}
  \end{align}
  which avoids the direct evaluation of the position operator in periodic
  boundary conditions~\cite{resta_macroscopic_1994}. 

  Finally, we reabsorb $\rho^{\Ea}_{\rm a,inter}(\w)$
  into the passive-subspace part of the density operator via a
  redefinition of the first-order wavefunctions, \beqref{eq:rhoE_full}, 
  obtaining a physically meaningful separation of ${\rho^{\Ea}(\w)=\rho^{\Ea}_{\rm intra}(\w)+\rho^{\Ea}_{\rm inter}(\w)}$ into intraband and interband parts, where the latter 
  is defined as ${\rho^{\Ea}_{\rm inter}(\w)=\rho^{\Ea}_{\rm a,inter}(\w)+\rho^{\Ea}_{\rm p}(\w)}$.

  By doing so, we naturally recover the textbook separation~\cite{allen_chapter_2006,resta_drude_2018} of
  the optical conductivity into a Drude and a regular part,
  %
  %It mediates the Drude contribution to the conductivit
  %is the Drude weight, so that the conductivity separates into Drude and
  %interband parts,
  \begin{align}
      \sigma^{\rm el}_{\alpha\beta}(\w) =
      \frac{D_{\alpha\beta}}{-i\w \pi}
      + \sigma^{\rm el,bound}_{\alpha\beta}(\w),
      \label{eq:sigma_decomp} 
  \end{align} 
  consistent with Ref.~\cite{marchese_born_2024}, Eq.~(45).  
  Applying Eq.~\eqref{eq:H_A_to_E} once more recovers the expected form of
  the clamped-ion dielectric tensor, Eq.~\eqref{eq:chi_el}, which in
  systems with time-reversal symmetry reduces to Eq.~(8) of
  Ref.~\cite{binci_first-principles_2021}. The additional term that
  vanishes under time-reversal symmetry is itself well known as
  the intrinsic anomalous Hall conductivity,
  \begin{align}
      \sigma^{\rm Hall}_{\alpha\beta} = -\frac{e^2}{\hbar}
      \int[d{\bf k}] \sum_l f_{\mathbf{k}l}\,
      2\,{\rm Im}\!\!\sum_{m \neq l}
      \frac{\bigl(H^{k_\alpha}\bigr)_{\mathbf{k}lm}
      \bigl(H^{k_\beta}\bigr)_{\mathbf{k}ml}}
      {\bigl(\varepsilon_{\mathbf{k}l} - \varepsilon_{\mathbf{k}m}\bigr)^2},
      \label{eq:hall_cond}
  \end{align}
  that is, the Brillouin-zone integral of the Berry
  curvature~\cite{thouless_quantized_1982,xiao_berry_2010}.
  % which vanishes in presence of time-reversal symmetry.
  % \CD{A bit confused here. You are saying the offdiagonal part vanishes? But the diagonal part (longitudinal conductivity) does not vanish, no?}

  % \subsubsection*{Implementation}

  %For the numerical evaluation of $\sigma^{\rm el,bound}_{\alpha\beta}(\w)$
  %and the electric-field perturbation,  The
  %electric-field matrix elements are related to the momentum matrix
  %elements via the Heisenberg equation of motion,

  %The passive-subspace perturbed
  %states $\ket{\psi^{\Ea}_{\mathbf{k}l}(\w)}_{\rm p}$ are
  %obtained from the Sternheimer equation~\eqref{eq:stern} evaluated in
  %the electric-field gauge.

  %The applied operator $\ket{i\psi^{k_\alpha}_{\mathbf{k}l}}=H^{\Ea}_{\rm bare}\ket{\mathbf{k}l}$, required as
  %the source term, is obtained as the 
  %while the active contribution is given by the explicit Kubo sum
  %\begin{align}
  %    \ket{\psi^{k_\alpha}_{\mathbf{k}l}}_{\rm a} =
  %    \sum_{m=1}^{M}\theta_{\mathbf{k}lm}\,\bar{F}_{\mathbf{k}lm}\,
  %    \bigl(H^{k_\alpha}_{\rm bare}\bigr)_{\mathbf{k}ml}
  %    \ket{\mathbf{k}m}.
  %\end{align}
  %Finally, 

  \begin{table*}[!]
\begin{tabular}{c|ccH|cccc}
$\omega+i\eta$ (eV) & $\sum_s Z_s(\w)$ & $-\w^2\chi^{\rm el}(\w)$ & Validation & $D$ & $-\w^2\chi^{\rm el, bound}(\w)$ & $\sum_s Z_s^{\rm nl}(\w)$\\
\midrule
\multicolumn{7}{c}{$\delta \rho=0$}                                                                                                                    \\
\midrule
$0.0+i0^{+}$ & $0.0003$ & $0.0$ & $0.0003$ & $0.0$ & $0.0$ & $0.0$\\
$0.5+i0^{+}$ & $-0.0467$ & $-0.0470$ & $0.0003$ & $0.0$ & $-0.0470$ & $0.0077$\\
$1.0+i0^{+}$ & $-0.1933$ & $-0.1936$ & $0.0003$ & $0.0$ & $-0.1936$ & $0.0332$\\
$1.5+i0^{+}$ & $-0.4775$ & $-0.4778$ & $0.0003$ & $0.0$ & $-0.4778$ & $0.0958$\\
      \midrule
      \multicolumn{7}{c}{$\delta \rho = -0.01$}                                                                                                                \\
      \midrule
$0.0+i0^{+}$ & $0.0396$ & $0.0394$ & $0.0002$ & $0.0394$ & $0.0000$ & $-0.0140$\\
$0.5+i0^{+}$ & $-0.0059$ & $-0.0061$ & $0.0002$ & $0.0394$ & $-0.0455$ & $-0.0072$\\
$1.0+i0^{+}$ & $-0.1460$ & $-0.1462$ & $0.0002$ & $0.0394$ & $-0.1856$ & $0.0148$\\
$1.5+i0^{+}$ & $-0.3967$ & $-0.3969$ & $0.0002$ & $0.0394$ & $-0.4363$ & $0.0587$\\
      \midrule
$0.0+i0.27$ & $0.0529$ & $0.0527$ & $0.0002$ & $0.0394$ & $0.0134$ & $-0.0160$\\
$0.5+i0.27$ & $0.0080-i0.0496$ & $0.0078-i0.0496$   & $0.0002$ & $0.0394$ & $-0.0316-i0.0496$ & $-0.0094+i0.0075$\\
$1.0+i0.27$ & $-0.1301-i0.1034$ & $-0.1303-i0.1034$ & $0.0002$ & $0.0394$ & $-0.1697-i0.1034$ & $0.0117+i0.0166$\\
$1.5+i0.27$ & $-0.3737-i0.1705$ & $-0.3739-i0.1705$ & $0.0002$ & $0.0394$ & $-0.4133-i0.1705$ & $0.0523+i0.0317$\\
\end{tabular}
    \caption{Validation of the generalized acoustic sum rule \beqref{eq:sumrule} 
  for BaSnO$_3$ at different doping levels and frequencies (for readability, conversion constants are omitted in the column headers). The left-hand side 
  $\sum_s Z_s(\w)$ and right-hand side $-\w^2\chi(\w)$ of Eq.~\eqref{eq:sumrule} 
  agree to within $3\times10^{-4}$, matching the accuracy of static DFPT 
  calculations in insulators and confirming the internal consistency of our implementation. 
  Individual contributions (Drude weight, interband susceptibility, and non-local 
  pseudopotential correction  \beqref{eq:bec_nl_pp}) are decomposed in the rightmost columns.
  In the calculations, the infinitesimal imaginary part $0^{+}$ (first column) is set to $10^{-8}$~Ry. }
    \label{tab:sumrule}
  \end{table*}

  \subsection{Traveling pseudopotential correction}\label{sec:travel_pp} 

  When testing the generalized acoustic sum rule Eq.~\eqref{eq:sumrule}
  within a pseudopotential framework, particular care must be taken to
  account for the non-local component of the ionic pseudopotential
  $V^{nl}_{\mathbf{k}s}$. As shown in Ref.~\cite{stengel_rototranslational_2026}, the motion of an ion induces
  an additional contribution to the deformation potential beyond the
  standard local term, arising from the dependence of the non-local
  pseudopotential on the ionic momentum. The corrected deformation potential
  reads
  \begin{align}
    V^{(R_{s\alpha})}(\w) =
    V^{(R_{s\alpha})}_{\rm loc}(\w)
    - i\hbar \w\, v^{nl}_{{\bf k} s\alpha},
    \label{eq:def_pot_corr}
  \end{align}
  where $v^{nl}_{\mathbf{k},s\alpha} = \frac{m_e}{\hbar^2}\partial V^{\rm ps}_{\mathbf{k}s}/
    \partial k_\alpha$ is the non-local contribution to the electron velocity
  operator  associated with ion $s$ and 
  $V^{\rm ps}_{\bf k}$ is the  cell-periodic pseudopotential operator
  defined in \beqref{eq:Vps_def}.
  % whose $k$-component in real-space reads
  % \begin{align}
  %   \hat{V}^{\rm ps}_{\bf k}     & =                 
  %   \int d{\bf r} d{\bf r}' 
  %   V^{\rm ps}({\bf r}',{\bf r}) e^{i {\bf k}\cdot({\bf r}'-{\bf r})}\op{\bf r}{\bf r'}  \nonumber                                                                        \\
  %   V^{\rm ps}({\bf r}',{\bf r}) & = \sum_{sL} e^{-i(m_e \dot{\bf R}^{L}_{s}/\hbar)\cdot({\bf r}'-{\bf r})} V^{\rm ps}_{s}({\bf r}-{\bf R}^{L}_{s},{\bf r}'-{\bf R}^{L}_{s}).
  %   \label{eq:Vps_def}
  % \end{align}

  This correction generates an additional contribution to the Born
  effective charges,
  \begin{align}
    Z^{\rm nl}_{s\alpha\beta}(\w) = \frac{i\hbar \w }{e}
    \int [d{\bf k}]\sum_{lm}
    F_{\mathbf{k}lm}(\w)\,
    \bigl( v^{nl}_{s\alpha}\bigr)_{\mathbf{k}lm}\,
    \bigl(H_{\rm scf}^{\E_\beta}(\w)\bigr)_{\mathbf{k}ml},
    \label{eq:bec_nl_pp}
  \end{align}
  where $H_{\rm scf}^{\E_\beta}(\w)$ is the self-consistent perturbation induced
  by an electric field along $\beta$.
      The diagonal ($l=m$) terms of Eq.~\eqref{eq:bec_nl_pp} are ill-defined as written,
since $F_{\mathbf{k}ll}(\w)=0$ at $\w\neq0$ while $(H^{\E_\beta}_{\rm bare})_{\mathbf{k}ll}$
diverges. They are resolved by combining Eq.~\eqref{eq:A_to_E} with an
algebraic manipulation that leads to 
% the identity
% $\hbar\w/[\Delta(\Delta+\hbar\w)]=1/\Delta-1/(\Delta+\hbar\w)$,
% $\Delta=\e_{\mathbf{k}l}-\e_{\mathbf{k}m}$, which yields
% \begin{align}
%   \hbar\w\,F_{\mathbf{k}lm}(\w)\,\bigl(H^{\E_\beta}_{\rm bare}\bigr)_{\mathbf{k}ml}
%   = ie\,\bigl(H^{k_\beta}_{\rm bare}\bigr)_{\mathbf{k}ml}
%   \bigl[\bar F_{\mathbf{k}lm}-F_{\mathbf{k}lm}(\w)\bigr],
%   \label{eq:wF_identity}
% \end{align}
% which is regular for $l=m$, where $\bar F_{\mathbf{k}ll}=f'_{\mathbf{k}l}$ and
% $F_{\mathbf{k}ll}(\w)=0$. The self-consistent part of the perturbation does not
% contribute to the diagonal terms, since $\w F_{\mathbf{k}ll}(\w)=0$.
% We thus 
separate the interband contribution
\begin{align}
Z^{nl,{\rm bound}}_{s\alpha\beta}(\w) &=
\frac{i\hbar\w}{e}\int [d{\bf k}]\sum_{\substack{lm \\ l\ne m}}
F_{\mathbf{k}lm}(\w)\,\nonumber\\&
\times \bigl(v^{nl}_{s\alpha}\bigr)_{\mathbf{k}lm}\,
\bigl(H^{\E_\beta}_{\rm scf}(\w)\bigr)_{\mathbf{k}ml},
  \label{eq:bec_nl_pp_inter}
\end{align}
from the intraband one,
\begin{align}
  Z^{nl,{\rm intra}}_{s\alpha\beta} =
  -\int [d{\bf k}]\sum_{l}
  f'_{\mathbf{k}l}\,
  \bigl(v^{nl}_{s\alpha}\bigr)_{\mathbf{k}ll}\,
  \bigl(H^{k_\beta}_{\rm bare}\bigr)_{\mathbf{k}ll},
  \label{eq:bec_nl_pp_intra}
\end{align}
which is real, frequency independent, and confined to the Fermi surface by $f'_{\mathbf{k}l}$.

 The intraband term is thus only present in metals while
  the explicit factor of
  $\w$ in Eq.~\eqref{eq:bec_nl_pp_inter} means that the bound correction vanishes in
  the static limit, so that both are absent from standard insulating DFPT calculations. We implement this correction, to our knowledge for the first time in a conducting system at finite frequency, and quantify its role in
  satisfying the sum rule~\beqref{eq:sumrule}. Our results complete the validation of the generalized
  acoustic sum rule across all regimes: previously it had been verified
  only at zero frequency in metals~\cite{dreyer_nonadiabatic_2022,marchese_born_2024} 
  and at finite frequency in  insulators~\cite{stengel_rototranslational_2026}.

  \subsection{Polynomial interpolation strategy}\label{sec:interp}

  The computational cost of evaluating the full frequency dependence of the
  response functions entering $\mathbb{S}(\w)$ at many frequencies and
  doping levels is substantial. The framework presented here is particularly
  well suited to an efficient interpolation strategy that reduces this cost
  while preserving accuracy.

  The key observation is that, although $\mathbb{S}^{-1}(\w)$ diverges at the
  coupled plasmon-phonon resonances and cannot be interpolated directly,
  $\mathbb{S}(\w)$ is a smooth function of frequency, whose determinant crosses zero
  at the poles, and is therefore well approximated by a low-order
  polynomial or spline. Furthermore, the individual ingredients of $\mathbb{S}(\w)$
  ---the Born effective charges, the interatomic force constants, and the
  interband dielectric response---are only weakly dependent on the doping
  level, far from significant electronic resonances, so that a single polynomial
  expansion in both frequency and doping captures their behavior across
  the insulator-to-metal transition at modest computational cost.

  Full calculations at a small number of judiciously chosen frequencies
  serve two complementary purposes: they provide the expansion coefficients
  for the interpolation, and they allow the accuracy of the interpolated
  $\mathbb{S}(\w)$ to be tested pointwise against the exact result.
  The generalized acoustic sum rule Eq.~\eqref{eq:sumrule} provides an
  additional global check: since it must be satisfied at every frequency,
  any systematic error introduced by the interpolation manifests as a
  violation of the sum rule, enabling rapid identification of frequency
  regions where additional exact calculations are needed. Together, these
  ingredients constitute a computationally efficient strategy with a built-in consistency check for mapping the coupled plasmon-phonon spectrum across the
  insulator-to-metal transition.

  \section{Low-energy spectrum of BaSnO\texorpdfstring{$_3$}{3}}
  \label{sec:appl_res}

  \subsection{Material and computational details}\label{sec:bso_details}

  BaSnO$_3$ (BSO) is a cubic perovskite oxide whose rare combination of a
  wide optical gap and high electron mobility upon La substitution makes it
  a leading transparent conducting oxide. It is also an ideal testbed for
  the present framework, since varying the La concentration sweeps the
  plasma frequency across the polar phonon frequencies, traversing all
  three regimes of the plasmon-phonon interplay within a single material.
  % Previous first-principles studies of BSO have focused primarily on its
  % electronic structure and transport
  % properties~\cite{monserrat_phonon-assisted_2018}; 
  % here 
  Here we exploit the extended dynamical matrix framework to characterize
  the coupled plasmon-phonon spectrum across the insulator-to-metal
  transition, providing both a benchmark for the method and a detailed
  picture of the low-energy excitations relevant to far-infrared
  spectroscopy.

  All calculations are performed within DFT as implemented in a custom
  version of the Quantum~ESPRESSO package~\cite{giannozzi_quantum_2009,
    giannozzi_advanced_2017}, using the local-density approximation with the Perdew--Wang
  parametrization of the correlation energy~\cite{perdew_accurate_1992} and
  scalar-relativistic optimized norm-conserving Vanderbilt
  pseudopotentials~\cite{hamann_optimized_2013} from the PseudoDojo table~\cite{van_setten_pseudodojo_2018} (Ba $5s^25p^66s^2$,
  Sn $4s^24p^64d^{10}5s^25p^2$, and O $2s^22p^4$ in valence), with
  plane-wave cutoffs of 100~Ry for the wave functions and 400~Ry for the
  charge density. For all doping levels we use the theoretical lattice parameter of
  the undoped crystal, $a = 4.087$~\AA\ (experiment: 4.116~\AA).
  The Brillouin zone is sampled on a shifted $32^3$ Monkhorst--Pack grid,
  with a Gaussian smearing of 0.012~Ry in the doped cases.
% \g{[PR: referee: smearing 0.012~Ry (163 meV) $\gg E_F$ at $10^{-5}$ e/cell and the $32^3$ grid does not resolve $k_F$; a convergence check (or a sentence that the low-doping points are checked against the EPIq $144^3$, $\sigma=0.0019$~Ry curve) is still missing.]}  
Doping is modeled by adding $|\delta\rho_{\rm dop}| = 10^{-5}$, $10^{-4}$,
  $10^{-3}$, $10^{-2}$, and $3\times10^{-2}$ electrons per cell with a
  compensating jellium background. The frequency-dependent response
  functions entering $\mathbb{S}(\w)$ are computed by TD-DFPT
  at a set of frequencies and doping levels, and then
  extended across the spectrum by the polynomial interpolation strategy
  described in Sec.~\ref{sec:interp}.

  In certain applications it may be advantageous to evaluate different
  ingredients of $\mathbb{S}(\w)$ with different numerical strategies,
  at the cost of forfeiting the exact equal-footing treatment required for
  sum-rule validation. A practically important case is the Drude weight,
  which converges slowly with Brillouin-zone sampling and is therefore
  particularly well suited to Wannier-function interpolation of the band
  curvature onto a dense $\mathbf{k}$-grid. We follow this approach for
  the plasma frequency results of \ref{fig:bso_plasma}, using the
  \textsc{EPIq} package~\cite{marini_epiq_2024}.
  Wannier model (EPIq curve of \ref{fig:bso_plasma}): ten maximally localized
  Wannier functions (Wannier90), O $2p$ and Sn $5s$ initial projections, from the undoped
  ground state (fixed occupations) on an unshifted $12^3$ grid. 
  % Bands 1--16 (semicore,
  % O $2s$, Ba $5p$) excluded; the remaining 14 bands disentangled with frozen window from
  % $-9.41$ to $+4.19$~eV and outer window up to $+9.19$~eV, relative to the valence-band
  % maximum (11.165~eV; conduction-band minimum 12.287~eV), i.e.\ O $2p$ manifold plus the
  % bottom of the Sn $5s$ band frozen. Total spread 15.99~\AA$^2$. 
  Interpolation on a
  shifted $144^3$ grid, rigid-band doping by raising $E_F$, Gaussian smearing
  0.0019~Ry. 
  % An $8^3$ Wannier grid gives the same Drude weight within 1\%; a $4^3$
  % grid overestimates it by $\sim$16\%.
  % where to put this figure.}

  \subsection{Validation of the generalized acoustic sum rule}
  \label{sec:bso_sumrule}

  \begin{figure}[h]
    \centering
    \includegraphics[width=0.99\linewidth]{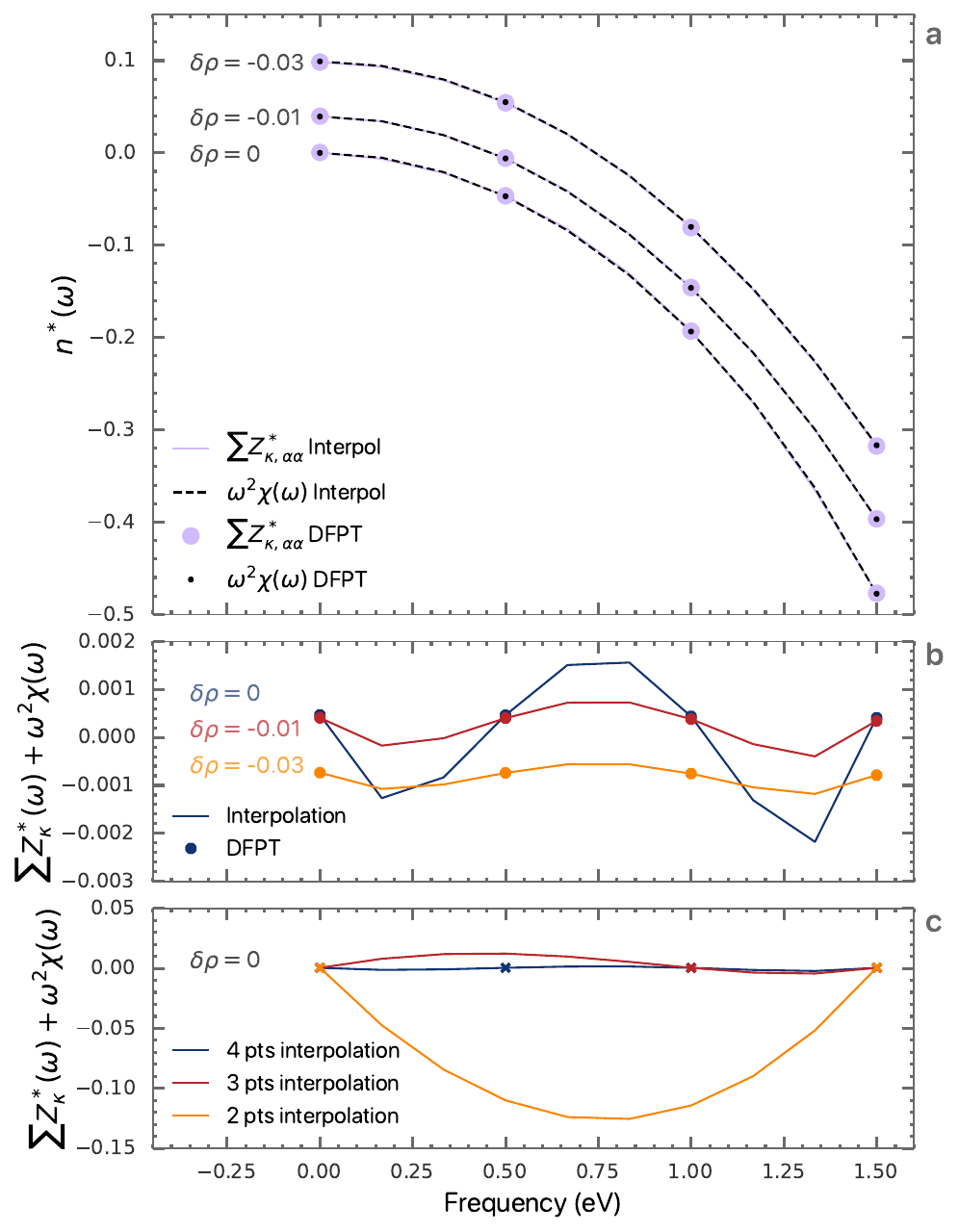}
    \caption{Validation of the generalized acoustic sum rule
      Eq.~\eqref{eq:sumrule} in BaSnO$_3$ as a function of frequency,
      for several doping levels $\delta\rho$. In panel a, solid lines show the left-hand
      side $\sum_s Z_{s,xx}(\w)$ and the right-hand side
$D - (m_e\Omega/e^2)\,\w^2 \chi^{\rm el,bound}_{xx}(\w)$ evaluated from
      TD-DFPT calculations (markers) and from interpolation (curves).
      Panel b shows the corresponding sum-rule residual. The
      residual error remains below $10^{-3}$ across all doping levels and
      frequencies, comparable to the accuracy achieved in the static
      adiabatic case for insulators, and does not increase significantly
      upon interpolation. The absolute residual is similar across 
      the insulator-to-metal transition, from the undoped ($\delta\rho = 0$) to 
      the most heavily doped ($\delta\rho =-0.03$) case.
      Panel c illustrates the role of the sum rule in benchmarking the interpolation quality,
      showing how the error of the sum rule increases when the interpolation is performed over a reduced 
      number of frequencies, using 2, 3, or 4 TD-DFPT points.}
    \label{fig:sumrule}
  \end{figure}

  We begin by benchmarking the internal consistency of the method through
  the generalized acoustic sum rule Eq.~\eqref{eq:sumrule}, including
  the traveling pseudopotential correction of
  Ref.~\cite{stengel_rototranslational_2026}. Table~\ref{tab:sumrule}
  reports the results for several doping levels and frequencies, including
  points with a finite imaginary part. The non-local pseudopotential
  correction is found to contribute even in the zero-frequency limit in metals at the level of $\sim$35\% of the total Born
  effective charge sum, and larger at finite $\w$.

  \ref{fig:sumrule} shows the sum rule as a function of frequency
  for several doping levels. Panel (a) compares the two sides of
  Eq.~\eqref{eq:sumrule}, with
  markers denoting the TD-DFPT calculations and curves the polynomial
  interpolation. The corresponding residual, shown in panel (b), remains
  comparable to the error of standard static DFPT in insulators, across all 
  doping levels and frequencies, confirming
  that the finite-frequency generalization introduces no additional
  numerical error. The residual is equally small across the full range
  from the undoped ($\delta\rho = 0$) to the most heavily doped
  ($\delta\rho = -0.03$) case, demonstrating that the method maintains its
  accuracy throughout the insulator-to-metal transition. Notably, the
  interpolation itself introduces no significant degradation: the
  interpolated curves satisfy the sum rule to within the same bound as the
  exact points from which they are constructed. 
 
  Panel (c) illustrates the diagnostic value of the sum rule. Repeating
  the interpolation with a progressively reduced frequency grid --- using
  four, three, or two TD-DFPT points --- produces a systematic growth of the
  residual, which becomes apparent well before the interpolated response
  functions themselves show any obvious deficiency. The sum rule therefore 
  provides a sensitive and inexpensive criterion for deciding how many
  exact calculations are needed, and for identifying the frequency regions
  where they should be placed.

  \begin{figure}[h]
    \centering
    \includegraphics[width=0.95\linewidth]{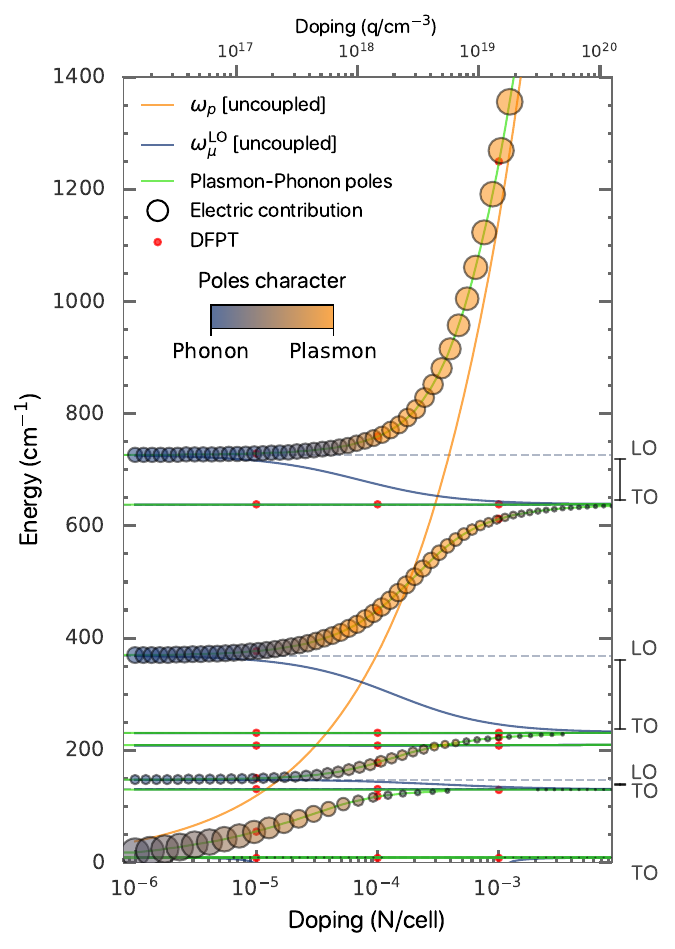}
    \caption{Plasmon-phonon resonances as a function of doping.  Dotted lines mark the LO and TO phonon frequencies in the adiabatic limit. Green solid lines report the coupled plasmon-phonon excitations. Solid lines represent the phonon (blue) and plasmon (yellow) resonances neglecting plasmon-phonon coupling. Red markers are results of uninterpolated DFPT calculations.
      The circles overlaying hybrid excitations characterize each mode, where the marker size represents the coupling with the electric field $c^{\E}_\lambda$ and its color depicts the plasmon character, quantified by $|\pdv{\omega_{\lambda}}{\delta\rho_{\rm dop}}|$. 
      }
    \label{fig:bso_disp}
  \end{figure}
  \subsection{Plasmon-phonon interplay across the insulator-to-metal
    transition}\label{sec:bso_coupling}

  With the accuracy of the method established, we turn to the physical
  characterization of the coupled plasmon-phonon spectrum of BSO. As the
  carrier density increases, the ratio $\omega_p/\omega_\mu$ between the
  plasma frequency and the polar phonon frequencies evolves through three
  qualitatively distinct regimes, illustrated in \ref{fig:bso_disp}.

  In the low-doping regime ($\omega_p \ll \omega_\mu$), the plasmon lies
  below the polar phonon frequencies. The electronic screening is only
  partial, internal electric fields are not fully suppressed, and
  longitudinal phonon modes retain a significant electric energy
  contribution that splits them above their transverse counterparts.

  At intermediate doping ($\omega_p \sim \omega_\mu$), the plasma frequency
  approaches one or more phonon frequencies and the two types of excitation
  hybridize strongly. The resulting eigenmodes are neither pure plasmons nor
  pure phonons but mixed plasmon-phonon hybrid modes whose character evolves
  rapidly with carrier density.

  In the high-doping metallic limit ($\omega_p \gg \omega_\mu$), the
  electronic screening completely suppresses the macroscopic electric field
  associated with polar phonons, so that the LO-TO splitting vanishes and
  longitudinal and transverse phonon frequencies become degenerate,
  $\omega^{\rm LO}_\mu = \omega^{\rm TO}_\mu$.

  \begin{figure}[!]
    \centering
    \includegraphics[width=\linewidth]{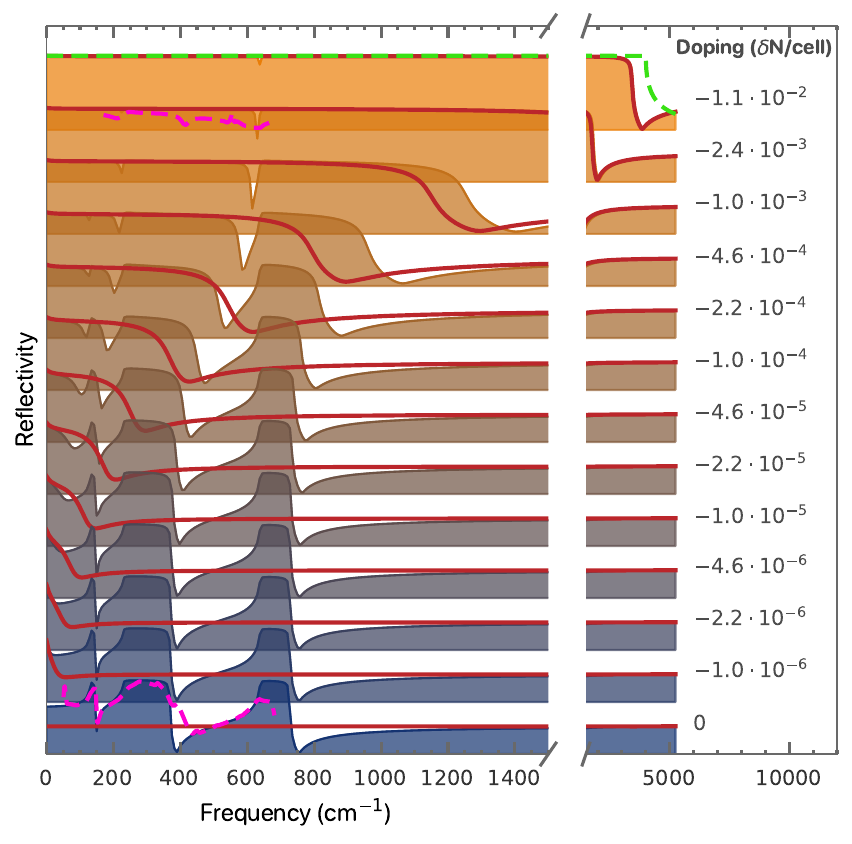}
    \caption{Reflectivity spectra of BaSnO$_3$ for increasing carrier density
      $\delta\rho$ (bottom to top). Shaded solid curves: full response,
      including the vibrational contribution to the dielectric function; red curves:
      purely electronic reflectivity at clamped ions. Dashed lines: reflectivity calculated from the experimental
      data Ref.~\cite{stanislavchuk_electronic_2012} (pink) and
      Ref.~\cite{james_allen_conduction_2016} (green).
      The vibrational contribution adds spectral features absent at clamped ions:
      a phonon dip at low doping, which progressively evolves into a
      plasmon-phonon hybrid structure as the carrier density increases.}
    \label{fig:bso_R}
  \end{figure}

  \begin{figure*}[!]
    \centering
    \includegraphics[width=0.85\linewidth]{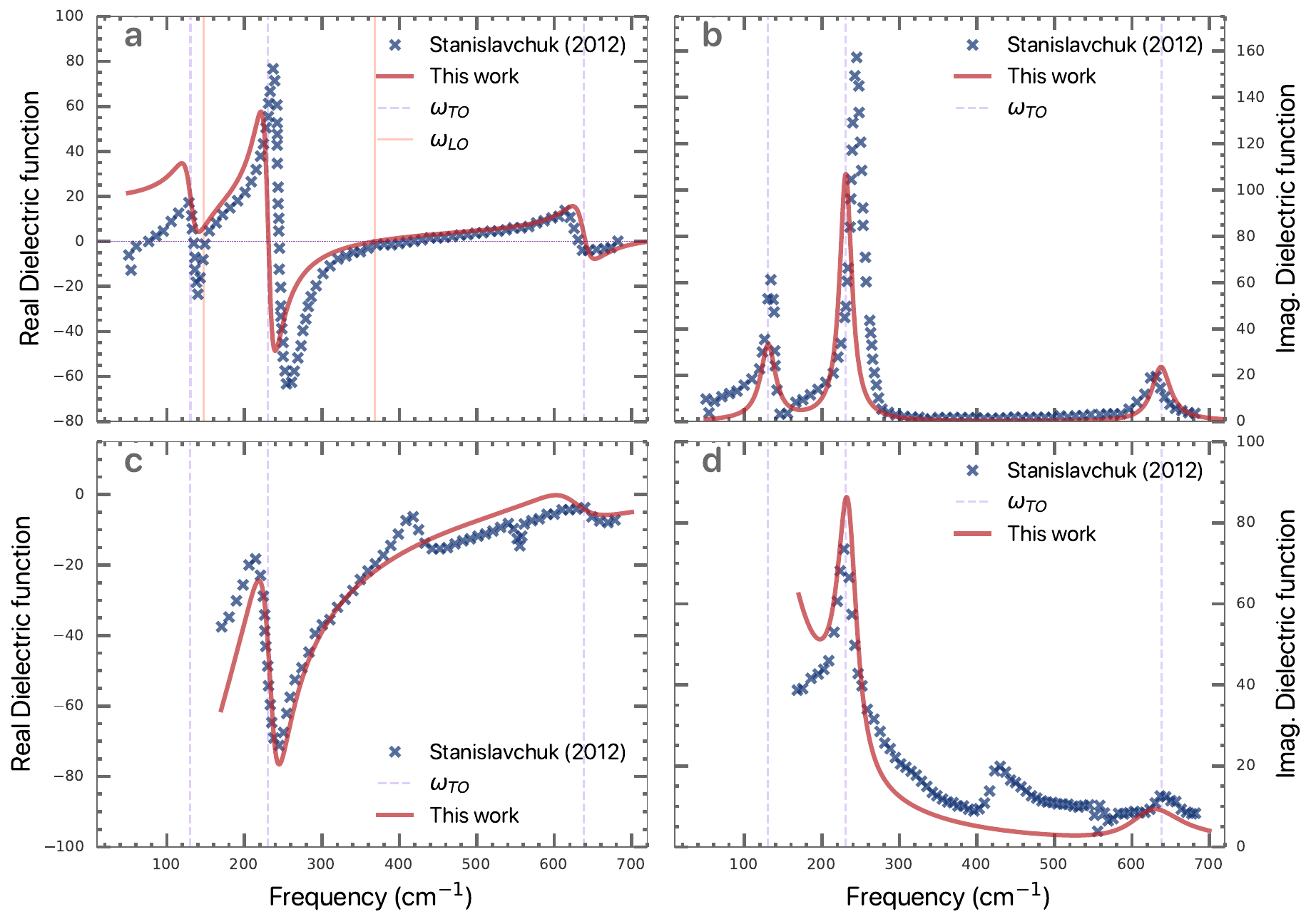}~%
    \caption{Far-infrared dielectric tensor in undoped (top) and doped (bottom) scenarios. The real and imaginary parts of the dielectric function are shown at zero doping in panels \textbf{a,b} and in the doped case in panels \textbf{c,d}. Theoretical calculations (solid curves) are compared with the experimental data~\cite{stanislavchuk_electronic_2012}.}
    \label{fig:bso_dielec}
  \end{figure*}

  The nature of each hybrid resonance can be further characterized by
  recasting the secular problem of Eq.~\eqref{eq:eom} as a
  frequency-dependent eigenvalue problem,
  \begin{align}
      \det\bigl[\mathbb{K}(\w) - \w^2\bigr] = 0,
      \label{eq:secular}
  \end{align}
  where $\mathbb{K}(\w)$ is a matrix with homogeneous units, obtained by
  working with mass-scaled displacements $u_s = \sqrt{m_s}\,\delta R_s$ and
  a rescaled potential $\widetilde{A} = \sqrt{4\pi/\Omega}\,A$.
  The squared resonance frequencies $\omega_\lambda^2$ are the eigenvalues
  of $\mathbb{K}(\omega_\lambda)$, determined self-consistently, and the
  hybrid nature of each mode is encoded in the corresponding eigenvector
  $\bm{\mathcal{V}}_\lambda = (\mathcal{V}^{\E}_\lambda,
  \{\mathcal{V}^\mu_\lambda\})$. We quantify the electric character of
  mode $\lambda$ as
  \begin{align}
      c^{\E}_\lambda =
      \frac{|\mathcal{V}^{\E}_\lambda|}{|\bm{\mathcal{V}}_\lambda|},
      \label{eq:el_component}
  \end{align}
  which ranges from $c^{\E}_\lambda = 0$ for a purely transverse, nonpolar
  mode to $c^{\E}_\lambda = 1$ for a plasmon in the metallic regime; in
  \ref{fig:bso_disp} it is encoded in the size of the circles.
  As a complementary descriptor, we define the plasmonic character of a
  mode as the sensitivity of its frequency to the carrier density,
  $|\partial\omega_\lambda/\partial\delta\rho_{\rm dop}|$, encoded in the
  color of the circles in \ref{fig:bso_disp}. 

  The framework allows a direct comparison of approximation levels. The
  static treatment fails qualitatively: the off-diagonal terms of $\mathbb{S}$ vanish, 
  so that the avoided-crossing
  hybridization is entirely absent (blue and yellow solid curves). 
  The pseudo-static approximation of Eq.~\eqref{eq:stat_approx} keeps the analytic Drude term and freezes the other response functions at zero frequency. It performs remarkably well: resonances stay within 1~cm$^{-1}$ of the full TD-DFPT result in the phonon region, including avoided crossings, and within 0.2\% (5~cm$^{-1}$) along the plasmon branch across the whole doping range. They are therefore not shown in \ref{fig:bso_disp}.
  Freezing in addition $\chi^{\rm el,bound}$, $C^0$ and $Z^*$ at their undoped static values, as in Ref.~\cite{lihm_plasmon-phonon_2024}, leaves phonon-like resonances unchanged within 1~cm$^{-1}$ but lowers the plasmon branch by up to 1\% (29~cm$^{-1}$ at $\delta\rho \approx -9\times10^{-3}$) through the doping dependence of interband screening.

  The accuracy of these approximations reflects the wide optical gap that makes BSO a
  valuable transparent conductor: in the absence of low-energy interband
  transitions, $\chi^{\rm el,bound}(\w)$, $Z_{s\alpha\beta}(\w)$ and
  $C^0_{ss'\alpha\beta}(\w)$ are all essentially flat across the phonon and
  plasmon energy range. Narrow-gap semiconductors, semimetals and bad
  metals, where electronic transitions occur at comparable energies, are
  expected to behave differently, since their screening acquires a pronounced
  frequency dependence already at the vibrational scale. Being valid by
  construction across all such regimes, the present framework provides the
  benchmark needed to establish when the cheaper approximation is
  justified.

  \ref{fig:bso_R} shows the theoretical reflectivity spectra as a
  function of doping, obtained by evaluating the  Fresnel formula for normal incidence with the dielectric function
  from Eq.~\eqref{eq:diel_tot}. Comparison with the clamped-ion spectra
  (solid red curves) reveals two effects of the ionic contribution: the
  appearance of additional absorption features at low energy associated
  with polar phonon modes, and a significant renormalization of the plasma
  edge frequency due to the plasmon-phonon coupling discussed in
  Sec.~\ref{sec:pla_renorm}.

\subsection{Comparison with experiment}\label{sec:bso_exp}

% \add{Experimental information on the longitudinal low-energy excitations of
% this system is limited to the LO frequencies of the undoped crystal
% (from $-\Im[\epsilon^{-1}(\w)]$ and Raman scattering)~\cite{stanislavchuk_electronic_2012}
% and to the plasma resonance of a doped film~\cite{james_allen_conduction_2016},
% so a mode-by-mode comparison of the coupled spectrum across doping is not possible.}
% {We are not aware of experimental measurements of the longitudinal
% low-energy excitations of this system, so a direct comparison with
% experiment is not possible.}
% As shown in Sec.~\ref{sec:extD}, however, the longitudinal plasmon-phonon
% spectrum is uniquely determined by the transverse (short-circuit)
% dielectric function; comparing our calculated dielectric function with
% experimental infrared spectra therefore provides a valuable
% benchmark.

% The far-infrared dielectric functions of undoped BSO and of La-doped
% Ba$_{0.97}$La$_{0.03}$SnO$_3$ (BLSO) were measured at room temperature by
% spectroscopic ellipsometry in Ref.~\cite{stanislavchuk_electronic_2012}.
Two experiments probe the low-energy excitations of this system: the plasma resonance of a doped film~\cite{james_allen_conduction_2016}, and the dielectric function from spectroscopic ellipsometry~\cite{stanislavchuk_electronic_2012}.
As shown in Sec.~\ref{sec:extD}, the longitudinal plasmon-phonon spectrum is uniquely determined by the transverse (short-circuit) dielectric function. Comparing our calculated dielectric function with experimental infrared spectra is therefore a valuable benchmark.

Ref.~\cite{stanislavchuk_electronic_2012} measured the far-infrared dielectric functions of undoped BSO and of La-doped Ba$_{0.97}$La$_{0.03}$SnO$_3$ (BLSO) at room temperature. 
These measurements give access to both the vibrational and the electronic
contributions to the dielectric function, in the insulating and in the
doped crystal.
By fitting the spectra with a set of Lorentz oscillators plus, for BLSO, a
Drude term, the authors extracted the frequencies of the three
infrared-active TO phonons (135, 244.5, and 628~cm$^{-1}$), the
high-frequency dielectric constant ($\epsilon_\infty = 3.3$), and the
plasma frequency of the free carriers
($\omega_p=\sqrt{4 D / \epsilon_\infty} = 1156 \pm 110$~cm$^{-1}$).
From the inverse of the fitted dielectric function, they also obtained the longitudinal
frequencies as the maxima of $-\Im[\epsilon^{-1}(\w)]$, two of which (LO$_2$, LO$_3$) were confirmed by Raman scattering.

Our calculated dielectric function of undoped BSO agrees well with these
measurements [red curves in \ref{fig:bso_dielec}(a,b)].
% , and so do the
% fitted parameters, as summarized in Table~\ref{tab:bso_exp}.
The TO frequencies agree within 6\%, and the LO frequencies within 5\%,
with the exception of LO$_2$, which we underestimate by 12\%.
The underestimation of LO$_2$ has two origins, both related to the LDA.
First, the underestimated band gap makes the electronic dielectric constant
too large: we obtain $\epsilon_\infty = 4.77$, compared with
$\epsilon_\infty = 3.3$ from the far-infrared fit and
$\epsilon_1 \approx 4.3$ measured at 0.75~eV in the same
work~\cite{stanislavchuk_electronic_2012}, which reduces all LO-TO splittings.
Second, the total oscillator strength of the three polar modes, fixed by the
Born effective charges, agrees with experiment, but its distribution
among the modes does not: the LDA eigenvectors assign too little of it to
mode~2 and too much to mode~3.
Since LO$_2$ is screened by both $\epsilon_\infty$ and the TO$_3$ oscillator,
it is the mode most affected, whereas for LO$_3$ the two errors partly
compensate.
Hybrid-functional (HSE) calculations correct both effects: they reduce
$\epsilon_\infty$ to 3.93, shift oscillator strength back to mode~2, and
place LO$_2$ at 422~cm$^{-1}$, in agreement with
experiment~\cite{monserrat_phonon-assisted_2018}.
% \begin{table}[h]
%   \caption{Infrared parameters of undoped BSO and doped BLSO: present
%     calculation vs.\ the ellipsometry fit of
%     Ref.~\cite{stanislavchuk_electronic_2012} and the plasma-resonance
%     measurement on a La-doped film of Ref.~\cite{james_allen_conduction_2016}.
%     Frequencies and scattering rates $\gamma$ in cm$^{-1}$.}
%   \label{tab:bso_exp}
%   \begin{ruledtabular}
%   \begin{tabular}{lcc}
%                        & Theory & Experiment \\
%     \hline
%     \multicolumn{3}{l}{Undoped BSO, Ref.~\cite{stanislavchuk_electronic_2012}} \\
%     $\omega_{\rm TO}$  & 138, 246, 648 & 135, 245, 628 \\
%     $\omega_{\rm LO}$  & 162, 374, 744 & 154, 421, 723 \\
%     $\gamma$           & 24, 12, 30\footnotemark[1] & 10, 18, 31 \\
%     $\epsilon_\infty$  & 4.77 & 3.3 (Far-IR), 4.3 (0.75 eV)\\
%     \hline
%     \multicolumn{3}{l}{Doped BLSO, Ref.~\cite{stanislavchuk_electronic_2012}} \\
%     $\sqrt{ 4D}$         & $\approx 1800$\footnotemark[1] & $2100 \pm 200$ \\
%     $\delta\rho_{\rm dop}$ & $\approx -5\times10^{-4}$\footnotemark[1] & $-0.03$\footnotemark[3] \\
%     \hline
%     \multicolumn{3}{l}{La-doped BSO, Ref.~\cite{james_allen_conduction_2016}} \\
%     $\delta\rho_{\rm dop}$ & $-0.011$%\footnotemark[4] 
%     & $-0.011$\\ %\footnotemark[5] \\
%     $\sqrt{ 4D}$         & $\approx 7600$ & $\approx 8200$\\%\footnotemark[6] \\
%   \end{tabular}
%   \end{ruledtabular}
%   \footnotetext[1]{Adjusted to the experimental spectra.}
%   \footnotetext[3]{Nominal La content.}
% \end{table}

We fitted a separate scattering rate to each phonon peak equal to 24, 18, 30 cm$^{-1}$. 
The authors of Ref.~\cite{stanislavchuk_electronic_2012} attribute the low-frequency trend, which our calculation does not reproduce, to diffraction and to diffuse scattering by the edges of the small crystals, which affect the spectra below about 150~cm$^{-1}$.

In the doped case [\ref{fig:bso_dielec}(c,d)],  we use  phonon-specific scattering rates of 24 and 100~cm$^{-1}$ for the vibrational part
and we include the Drude contribution as $-4D/(\w (\w+i\eta_{\rm Drude}))$, where the Drude weight is computed as a function of the doping level. We find as optimal values $\delta \rho=-5\times 10^{-4}$ and $\eta_{\rm Drude}=110$~cm$^{-1}$. 
The real and imaginary parts of the experimental dielectric function fix the carrier density and the
scattering rate, respectively, and a single pair reproduces both
[\ref{fig:bso_dielec}(c,d)]. 
The most visible discrepancy with the experimental data is an additional
peak at 427~cm$^{-1}$, which the authors of
Ref.~\cite{stanislavchuk_electronic_2012} assigned to M- or
R-point phonons made infrared-active by zone folding, as would result from
a local breaking of the cubic symmetry. Raman measurements also support this argument.

\begin{figure}[h]
  \centering
  \includegraphics[width=0.95\linewidth]{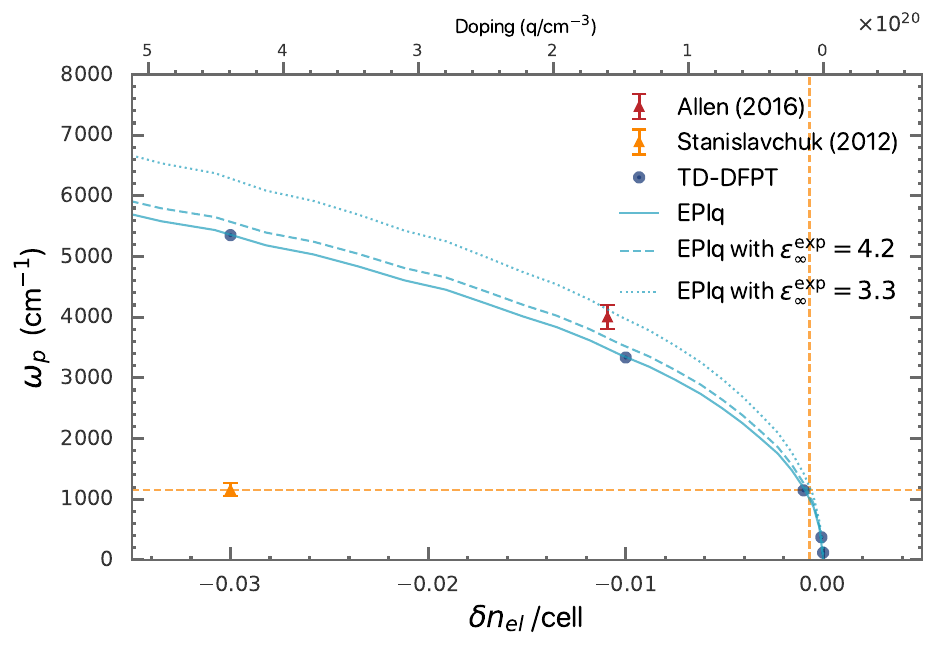}~%
  \caption{Plasma frequency as a function of doping level, from self-consistent calculations (circles, $N_k=32^3$, $\sigma_{\rm el}=0.012$~Ry), Wannier interpolation via EPIq~\cite{marini_epiq_2024} with the LDA ($\epsilon_\infty = 4.77$, solid curve) or the experimental ($\epsilon_\infty = 4.2$~\cite{james_allen_conduction_2016}, dashed curve and $\epsilon_\infty = 3.3$ \cite{stanislavchuk_electronic_2012}, dotted curve) dielectric constant ($N_k=144^3$, $\sigma_{\rm el}=0.0019$~Ry), and experimental estimates (triangles)~\cite{stanislavchuk_electronic_2012,james_allen_conduction_2016}.}
  \label{fig:bso_plasma}
\end{figure}

The theoretical result agrees with the experimental data and with the parameters fitted therein. This also holds for our Drude weight when screened with the $\epsilon_\infty = 3.3$ used in that fit, or, equivalently, when the Drude weights $4D$ are compared directly.
We calculate the relation between the plasma frequency and the
doping level (\ref{fig:bso_plasma}) and checked  the convergence
of the TD-DFPT result against a Wannier interpolation on an ultradense
mesh.
In this way we find  a carrier density of $\delta\rho\approx -5\times10^{-4}$
corresponding to such Drude weight.  
 A value that is about 60 times below the nominal doping $\delta\rho_{\rm dop} = -0.03$ reported in Ref.~\cite{stanislavchuk_electronic_2012}, where no transport characterization accompanies the ellipsometry data.
This discrepancy can be explained by incomplete dopant activation: compensation by defects, La segregation and incorporation at inactive sites are common in doped perovskite oxides.

A quantitative test of the doping dependence is provided by
Ref.~\cite{james_allen_conduction_2016}, where carrier density and plasma
frequency were measured on the same sample, a La-doped BSO film thick 
enough to behave as bulk.
The plasma resonance is observed at
4000 $\pm 200$~cm$^{-1}$ in p-polarized reflectivity at grazing incidence
for a  Hall density of $n = 1.6\times10^{20}$~cm$^{-3}$
($\delta\rho_{\rm dop} \approx -0.011$)
(red triangle in \ref{fig:bso_plasma}).
At the same doping, our calculation gives $\omega_p \approx 3480$~cm$^{-1}$, about 13\% lower.
About half of this difference comes from the overestimated LDA dielectric constant
($\epsilon_\infty = 4.77$ instead of $\epsilon^{\rm exp}_\infty = 4.2$, the value adopted in Ref.~\cite{james_allen_conduction_2016}); with the experimental $\epsilon_\infty$, our Drude
weight gives $\omega_p \approx 3710$~cm$^{-1}$, 7\% below the measured value.

% Moreover, at low doping our $\sqrt{4D}$ scales as that of free carriers with
% a band-edge effective mass $m^* \approx 0.20\,m_e$, in agreement with the
% value $m^* = 0.19 \pm 0.01\,m_e$ extracted in
% Ref.~\cite{james_allen_conduction_2016}.
This level of agreement supports both the present approach and the
description of La doping as free carriers added to the conduction band,
% {This level of agreement supports both the present approach and the
% jellium model adopted to describe La doping,}
while the comparison with
Ref.~\cite{stanislavchuk_electronic_2012} indicates that the nominal La
content is not a robust estimator of the effective carrier density.

  \section{Conclusion}\label{sec:concl}

  In this work we have presented a first-principles framework for the full \textit{ab initio} treatment of
  plasmon-phonon coupling at finite frequency within TD-DFPT. Minimization of the time-dependent Ehrenfest action
   Eq.~(\ref{eq:action}), yields the plasmon-phonon coupling as a secular problem defined over both lattice and electric degrees of freedom. This
  formulation treats phononic and plasmonic excitations on an equal footing, retaining the
  full frequency dependence of the interatomic force constants, Born effective charges,
  and electronic dielectric response of the doped crystal. It thereby lifts the
  approximations that limit the most advanced approaches available in the
  literature~\cite{lihm_plasmon-phonon_2024}: the plasmon-pole model, neglect of local
  fields, truncated sum over a finite manifold of rigidly doped bands, use of
  ingredients computed for the undoped system, and static limit.

    The Hessian $\mathbb{S}(\w)$ of the action, 
  together with its Legendre-transform counterpart $\mathbb{S}^{-1}(\w)$, provides a transparent
  physical picture of the role of boundary conditions and a rigorous connection to
  previously established results~\cite{born_dynamical_1954,
    cochran_dielectric_1962, pick_microscopic_1970}. In particular, the Cochran--Cowley
  long-range correction to the interatomic force constants and the LO-TO splitting emerge
  naturally from simple matrix algebra, with the electronic dielectric response evaluated
  at the exact finite frequency rather than in the static limit.

  The internal consistency of the approach enabled us to validate the generalized acoustic sum rule
  \cite{dreyer_nonadiabatic_2022}, to our knowledge for the first time at finite complex frequency and in
  conducting systems, including the traveling pseudopotential correction
  \cite{stengel_rototranslational_2026}. The sum rule is satisfied to an accuracy
  comparable with the static insulating case, confirming the internal consistency of the
  implementation. Furthermore, we showed that the sum rule provides a stringent
  benchmark for interpolation strategies: the error in the sum rule is
  sensitive to the quality of the frequency interpolation, making it a practical
  tool for validating the accuracy of approximate calculations.

  A further analytical result concerns the renormalization of the plasma frequency by the
  ionic lattice. Exploiting the connection between the acoustic sum rule and momentum
  conservation, we demonstrated that the acoustic phonon branches renormalize the plasma
  frequency through a correction expressible in terms of the ratio of the electronic
  effective mass ($n^* m_e$) to the total ionic mass. In this picture, the relevant inertia of the
  coupled electron-ion system is governed by their reduced mass, providing a transparent
  reinterpretation of the generalized acoustic sum rule as a statement about the center-of-mass
  motion of the neutral system.

  We applied the framework to BaSnO$_3$, a prototypical wide-gap perovskite oxide with
  high electron mobility upon La doping, characterizing its low-energy spectrum across the
  insulator-to-metal transition. The plasmon-phonon hybridization evolves through three
  distinct regimes as a function of doping: an unscreened polar regime at low carrier
  density, a strongly coupled hybrid regime near resonance, and a fully screened metallic
  regime at high doping. In this wide-gap material the pseudo-static approximation is accurate to 1~cm$^{-1}$ (phonons) and 0.2\% (plasmon), and freezing the ingredients at zero doping adds at most 1\%. The plasma frequency obtained from the Drude weight agrees to within 7\% with the plasma resonance measured at known carrier density~\cite{james_allen_conduction_2016}, once the experimental $\epsilon_\infty$ is used,
  supporting both the method and the free-carrier description of La doping. 

  Several natural extensions of the present work are envisaged. First, the inclusion of a
  finite wave vector $\mathbf{q}$ would allow access to the full plasmon dispersion and
  to Landau damping near the particle-hole continuum. Second, the generalization to
  two-dimensional materials, where the long-range Coulomb interaction is qualitatively
  modified, represents a timely and practically relevant direction. 
  % Finally, the 
  % complete frequency dependence achieved in this work provides direct
  % access to the imaginary part of the frequency-dependent response functions.

\begin{acknowledgments}
M.S. thanks the CCQ at the Flatiron Institute for hospitality during the early stages of this work.
M.S. and G.M. acknowledge support from the Spanish MCIN/AEI/10.13039/501100011033 under grant PID2023-152710NB-I00 and under a Severo Ochoa Excellence award to ICMAB-CSIC, Grant CEX2023-001263-S.
C.E.D.\ acknowledges support from the National Science Foundation under Grant No.~DMR-2237674.
The Flatiron Institute is a division of the Simons Foundation.
\end{acknowledgments}

  %%%%%%%%%%%%%%%%%%%%
  % \clearpage
  %%%%%%%%%%%%%%%%%%%%

  \bibliography{biblio}% Produces the bibliography via BibTeX.

  \end{document}